\documentclass[english,a4]{smfart}
\usepackage{amsmath,amssymb}%

\usepackage{graphicx}  

\usepackage{psfrag}

\usepackage{color}

\usepackage{epstopdf}
\usepackage{epsfig}
\usepackage{wrapfig}
\usepackage{a4}
\usepackage{amssymb,latexsym}
\usepackage[mathscr]{eucal}
\usepackage{cite}
\usepackage{url}

\DeclareFontFamily{OT1}{rsfs}{} \DeclareFontShape{OT1}{rsfs}{m}{n}{
<-7> rsfs5 <7-10> rsfs7 <10-> rsfs10}{}
\DeclareMathAlphabet{\mycal}{OT1}{rsfs}{m}{n}
\def\scri{{\mycal I}}%
\def\scrip{\scri^{+}}%
\def\scrp{{\mycal I}^{+}}%
\def\Scri{\scri}

\begin{document}

\frontmatter
\author [P.T.~Chru\'sciel]{Piotr T.~Chru\'sciel\thanks{The author is grateful to the Mittag-Leffler Institute, Djursholm, Sweden,
for financial support and hospitality during part of work on
this paper.}}
\address {LMPT,
F\'ed\'eration Denis Poisson, Tours; Mathematical Institute and
Hertford College, Oxford}
 \email {chrusciel@maths.ox.ac.uk}
\urladdr {www.phys.univ-tours.fr/$\sim$piotr}

%
%
\newcommand{\mnote}[1]{}%
%

\newcommand{\zk}{{\mathring k}}

\newcommand{\Spn}{S^+_0}
\newcommand{\Spz}{S^+_0}
\newcommand{\mcEp}{{\mcE^+}}
\newcommand{\mcEpz}{{\mcE^+_0}}
\newcommand{\mcEm}{{\mcE^-}}
\newcommand{\mcEmz}{{\mcE^-_0}}

\newcommand{\ii}{{\mathrm{i}}}

\newcommand{\jlcasug}[1]{{\color{blue}\mnote{\color{blue}{\bf jlca suggests!}
}}{\color{blue}{#1}}\color{black}\ }

\newcommand{\hfourg}{\hat \fourg}

\newcommand{\oft}{{outer future trapped}}
\newcommand{\wpt}{{weakly past trapped}}

\newcommand{\kk}[1]{}

\newcommand{\mexp}{\mbox{\rm exp}}

\newcommand{\zR}{\mathring R}
\newcommand{\zD}{\mathring D}
\newcommand{\zA}{\mathring A}
\newcommand{\mzh}{\mathring h}

\newcommand{\hypwithhat}{\hathyp}
\newcommand{\hathyp}{\,\,\widehat{\!\!\hyp}}

\newcommand{\Span}{\mathrm{Span}}

\newcommand{\odoc}{\overline{\doc}}
\newcommand{\ohyp}{\,\,\overline{\!\!\hyp}}
\newcommand{\pohyp}{\partial\ohyp}

\newcommand{\aregular}{{an {\regular}}}
\newcommand{\regular}{$I^+$--regular}

\newcommand{\hthreeg}{h}

\newcommand{\doce}{\doc_\epsilon}

\newcommand{\odocIp}{\overline{\doc}\cap I^+(\Mext)}
\newcommand{\odocup}{{\doc}\cup \mcH_0^+}
\newcommand{\llambda}{\lambda}

\newcommand{\ue}{u_{\epsilon}}
\newcommand{\uee}{u_{\epsilon,\eta}}

\newcommand{\bmcM}{\,\,\,\,\widetilde{\!\!\!\!\mcM}}
\newcommand{\bfourg}{\widetilde{\fourg}}

\newcommand{\eean}{\nonumber\end{eqnarray}}

\newcommand{\lp}{\ell}

\newcommand{\tSp}{\tilde S_p}
\newcommand{\Sp}{S_p}

\newcommand{\dgtcp}{\dgt}
\newcommand{\dgtc}{\dgt}

\newcommand{\id}{{\rm id}}

\newcommand{\Zd}{{\mcZ}_{\dgt}}

\newcommand{\mcHp}{{\mcH^+}}
\newcommand{\mcHpz}{{\mcH^+_0}}
\newcommand{\mcHm}{{\mcH^-}}
\newcommand{\mcHmz}{{\mcH^-_0}}

\newcommand{\zN}{\mathring N}

\newcommand{\mcY}{{\mycal Y}}
\newcommand{\mcX}{{\mycal X}}

\newcommand{\Nndg}{N_{\mbox{\scriptsize\rm  ndh}}}
\newcommand{\Ndg}{N_{\mbox{\scriptsize\rm  dh}}}
\newcommand{\Nndh}{N_{\mbox{\scriptsize\rm  ndh}}}
\newcommand{\Ndh}{N_{\mbox{\scriptsize\rm  dh}}}
\newcommand{\Naf}{N_{\mbox{\scriptsize\rm  AF}}}
\newcommand{\alpharate}{\lambda}
\newcommand{\zalpha}{\mathring \alpha}

\newcommand{\puncti}{a_i}
\newcommand{\ai}{\puncti}

\newcommand{\mcS}{{\mycal S}}

\newcommand{\nic}{}
\newcommand{\phypa}{\partial\mcS_{(a)} }
\newcommand{\Ya}{{}^{(a)}Y }
\newcommand{\Ja}{{}^{(a)}J }
\newcommand{\Oma}{{}^{(a)}\Omega }
\newcommand{\Xa}{{}^{(a)}X }
\newcommand{\Sa}{{}^{(a)}S }
\newcommand{\Qea}{{}^{(a)}Q^E}
\newcommand{\Qba}{{}^{(a)}Q^B}
\newcommand{\phiea}{{}^{(a)}\Phi^E}
\newcommand{\phiba}{{}^{(a)}\Phi^B}
\newcommand{\mcHa}{{}^{(a)}\mcH}

\newcommand{\hypo}{\,\,\mathring{\!\! \hyp}}
\newcommand{\ohypo}{\overline{\hypo}}
\newcommand{\hypot}{\,\,\mathring{\!\! \hyp_t}}
\newcommand{\hypoz}{\,\,\mathring{\!\! \hyp_0}}
\newcommand{\ohypoz}{\overline{\hypoz}}
\newcommand{\ohypot}{\overline{\hypot}}

\newcommand{\Kz}{K_\kl 0}

\newcommand{\fourge}{\fourg_\epsilon}

\newcommand{\Cp}{ \mcC^+ }
\newcommand{\Cpe}{ \mcC^+ _\epsilon}
\newcommand{\Ct}{ \mcC^+_t}
\newcommand{\Ctm}{ \mcC^+_{t_-}}
\newcommand{\hStmR}{\hat S_{t_-,R}}
\newcommand{\hStR}{\hat S_{t,R}}
\newcommand{\hSzR}{\hat S_{0,R}}
\newcommand{\hSts}{\hat S_{\tau,\sigma}}

\newcommand{\Sone}{\Sz}
\newcommand{\Sonep}{\Sz}
\newcommand{\Soneq}{S_{0,q}}
\newcommand{\Stp}{S_{t,p}}
\newcommand{\Stq}{S_{t,q}}
\newcommand{\Sz}{S_0}

\newcommand{\bw}{\bar w}
\newcommand{\bzeta}{\bar \zeta}

\newcommand{\zMtwo}{\mathring{M}{}^2}
\newcommand{\Mtwo}{{M}{}^2}
\newcommand{\bMtwo}{{\bar M}{}^2}
\newcommand{\hMtwo}{{\hat M}{}^2}

\newcommand{\hypext}{\hyp_{\mbox{\scriptsize \rm ext}}}
\newcommand{\Mtext}{\Sext}
\newcommand{\Mint}{\mcM_{\mbox{\scriptsize \rm int}}}
\newcommand{\mcMext}{\Mext}

\newcommand{\mcHN}{\mcN}
\newcommand{\mcNH}{\mcH}

\newcommand{\hS }{{\hat S }}
\newcommand{\hSq}{{\hat S_q}}
\newcommand{\hSp}{{\hat S_p}}
\newcommand{\hSpn}{{\hat S_{p_n}}}
\newcommand{\hSqn}{{\hat S_{q_n}}}
\newcommand{\mcA}{\mycal A}
\newcommand{\mcZ}{\mycal Z}

\newcommand{\zh}{{\,\,\widetilde{\!\!\mcZ}}}
\newcommand{\dgt}{{\mycal Z}_{\mbox{\scriptsize \rm dgt}}}

\newcommand{\kl}[1]{{(#1)}}

\newcommand{\Uone}{{\mathrm{U(1)}}}

\newcommand{\Sm}{\ensuremath{\Sigma_{-}}}
\newcommand{\Nt}{\ensuremath{N_{2}}}
\newcommand{\Nth}{\ensuremath{N_{3}}}

\newcommand{\jlca}[1]{\mnote{{\bf jlca:} #1}}

\newcommand{\jlcared}[1]{{\color{red}\mnote{{\color{red}{\bf jlca:}
#1} }}}
\newcommand{\jlcachange}[1]{{\color{blue}\mnote{\color{blue}{\bf jlca:}
changed on 18.IX.07}}{\color{blue}{#1}}\color{black}\ }

\newcommand{\jlcachangeb}[1]{{\color{blue}\mnote{\color{blue}{\bf jlca:}
changed on 19.IX.07}}{\color{blue}{#1}}\color{black}\ }
\newcommand{\jlcachangec}[1]{{\color{blue}\mnote{\color{blue}{\bf jlca:}
changed on 20.IX.07}}{\color{blue}{#1}}\color{black}\ }
\newcommand{\jlcachanged}[1]{{\color{blue}\mnote{\color{blue}{\bf jlca:}
changed after 9.X.07}}{\color{blue}{#1}}\color{black}\ }
\newcommand{\jlcachangeXIII}[1]{{\color{blue}\mnote{\color{blue}{\bf jlca:}
changed after 13.X.07}}{\color{blue}{#1}}\color{black}\ }

\newcommand{\jlcaadded}[1]{{\color{blue}\mnote{\color{blue}{\bf jlca:}
added on 20.IX.07}}{\color{blue}{#1}}\color{black}\ }
\newcommand{\sse }{\Longleftrightarrow}
\newcommand{\se}{\Rightarrow}
\newcommand{\uu}{{\bf u}}
\newcommand{\bbR}{\mathbb{R}}
\newcommand{\bbN}{\mathbb{N}}
\newcommand{\bbZ}{\mathbb{Z}}
\newcommand{\bbC}{\mathbb{C}}
\newcommand{\bbH}{\mathbb{H}}

\newcommand{\cof}{\operatorname{cof}}
\newcommand{\dive}{\operatorname{div}}
\newcommand{\curl}{\operatorname{curl}}
\newcommand{\grad}{\operatorname{grad}}
\newcommand{\rank}{\operatorname{rank}}
\def\G{{\mycal G}}
\def\scro{{\mycal O}}
\def\Doc{\doc}
\def\scrip{\scri^{+}}%
\def\scrp{{\mycal I}^{+}}%
\def\Scri{\scri}
\def\scra{{\mycal A}}
\def\Scra{\cup_{i}\scra_i}
\def\scru{{\mycal U}}
\def\scrw{{\mycal W}}
\def\scrv{{\mycal V}}
\def\scrs{{\mycal S}}
\def\rot{{\mycal R}}
\def\khor{{\mycal H}}
\def\e{\wedge}
\def\d{\partial}
\def\KK{\phi^K}
\def\K0{\phi^{K_0}}
\def\Kdot{\phi^{K}\cdot}
\def\X.{\phi^{X}\cdot}
\def\normK{W}

\newcommand{\hyphat}{\,\,\,\widehat{\!\!\!\hyp}}
\newcommand{\hmcM}{\,\,\,\widehat{\!\!\!\mcM}}

\newcommand{\newF}{\lambda}

\newcommand{\Int}{\operatorname{Int}}
\newcommand{\Tr}{\operatorname{Tr}}

\newcommand{\myuu}{u}

\newcommand{\oX}{\overline X}
\newcommand{\oY}{\overline Y}
\newcommand{\op}{\overline p}
\newcommand{\oq}{\overline q}

\newcommand{\hg}{{\hat g}}
\newcommand{\mcEh}{{{\mycal E}^+_\hyp}}

{\catcode `\@=11 \global\let\AddToReset=\@addtoreset}
\AddToReset{equation}{section}
\renewcommand{\theequation}{\thesection.\arabic{equation}}

\newcommand{\ptcKcite}[1]{.\cite{#1}}

\newcommand{\fourg}{{\mathfrak g }}

\newcommand{\refcite}[1]{\cite{#1}}
\newcommand{\levoca}[1]{\ptc{#1}}

\newcommand{\umac}{\gamma}%
\newcommand{\metrict}{\threeg }%

\newcommand{\mcN}{{\mycal N}}
\newcommand{\mcNX}{{\mycal N(X)}}

\newcommand{\cf}{cf.,}

\newcommand{\gcirc}{{\mathring{g}}_{ab}}
\let\a=\alpha\let\b=\beta \let\g=\gamma \let\d=\delta\let\lb=\lambda
\newcommand{\sts}{space-times}
\newcommand{\calM}{{\mcM}}
\newcommand{\calH}{{\mcH}}
\newcommand{\el}{e^{2\lb}}
\newcommand{\hysf}{hypersurface}
\newcommand{\OOmega}{\omega}

\newcommand{\nopcite}[1]{}

\newcommand{\hypM}{\hyp}

\newcommand{\gschw}{h_{\mbox{\scriptsize \rm Schw}}}
\newcommand{\gsch}{\gschw}

\newcommand{\calSJr}{{\mycal  S}_{(J_0,\rho_0)}}
\newcommand{\Ima}{\mbox{\rm Im}}
\newcommand{\Ker}{\mbox{\rm Ker}}
\newcommand{\sgstatic}{{strictly globally static{}}}
\newcommand{\gstatic}{{globally static{}}}
\newcommand{\riemg}{g}
\newcommand{\riemgz}{g_0}
\newcommand{\hhat}{\ghat}
\newcommand{\DeltaL}{\Delta_{\mathrm L}}
\newcommand{\ghat}{\gamma}
\newcommand{\hzhat}{\gamma_0}
\newcommand{\holder}{H\"older }
\newcommand{\del}{\partial}
\newcommand{\Mbar}{{\overline M}}
\newcommand{\gbar}{{\overline \riemg}}
\newcommand{\dm}{{\partial M}}
\newcommand{\dminfty}{{\partial_\infty M}}
\newcommand{\dmo}{{\partial_0 M}}
  {\renewcommand{\theenumi}{\roman{enumi}}
   \renewcommand{\labelenumi}{(\theenumi)}}

\newcommand{\bS}{{\overline \Sigma}}
\newcommand{\pS}{{\partial \Sigma}}
\newcommand{\tp}{\tau_p}
\newcommand{\hF}{\hat F}
\newcommand{\signX}{\mathrm{sign}X}
\newcommand{\mnu}{\nu}
\newcommand{\homega}{{\widehat{\Omega}}}%
\newcommand{\kerp}{\mathrm{Ker}_p\nabla X}%
\newcommand{\kerpi}{\mathrm{Inv}_p}%
\newcommand{\const}{\mathrm{const}}
\newcommand{\ml}{{M_{0,n-2\ell}}}
\newcommand{\mil}{{M_{\mathrm{iso},\ell}}}
\newcommand{\mtwo}{{M_{0,n-2}}}
\newcommand{\mfour}{{M_{0,n-4}}}
\newcommand{\sell}{\pi_\Sigma(\ml)}
\newcommand{\sfour}{\pi_\Sigma(\mfour)}
\newcommand{\stwo}{\pi_\Sigma(\mtwo)}
\newcommand{\mi}{{M_{0,n-2i}}}
\newcommand{\mj}{{M_{0,n-2j}}}
\newcommand{\htau}{{\hat \tau}}
\newcommand{\ttau}{{\tilde \tau}}
\newcommand{\zM}{{\mathring{M}}}
\newcommand{\ztau}{{\mathring{\tau}}}
\newcommand{\zSigma}{{\mathring{\Sigma}}}
\newcommand{\sings}{{\Sigma_{\mathrm{sing,iso}}}}
\newcommand{\miso}{{M_{\mathrm{iso}}}}
\newcommand{\psings}{{\partial\Sigma_{\mathrm{sing,iso}}}}
\newcommand{\sing}{{\Sigma_{\mathrm{sing}}}}
\newcommand{\singt}{{\Sigma_{\mathrm{sing},0}}}
\newcommand{\psingt}{{\partial\Sigma_{\mathrm{sing},0}}}
\newcommand{\Ein}{\operatorname{Ein}}
\newcommand{\hlambda}{\hat \lambda}
\newcommand{\mLX}{{\mcL_X}}
\newcommand{\mcE}{{\mycal E}}
\newcommand{\mcC}{{\mycal C}}
\newcommand{\mcD}{{\mycal D}}
\newcommand{\mcW}{{\mycal W}}
\newcommand{\lormet  }{{\frak g}}
\newcommand{\ApSwSw}{Appendix~\ref{SwSs}}
\newcommand{\abs}[1]{\left\vert#1\right\vert}
\newcommand{\norm}[1]{\left\Vert#1\right\Vert}
\newcommand{\M}{\EuScript M}
\newcommand{\Lie}{\EuScript L}
\newcommand{\nablash}{\nabla{\kern -.75 em
     \raise 1.5 true pt\hbox{{\bf/}}}\kern +.1 em}
\newcommand{\Deltash}{\Delta{\kern -.69 em
     \raise .2 true pt\hbox{{\bf/}}}\kern +.1 em}
\newcommand{\Rslash}{R{\kern -.60 em
     \raise 1.5 true pt\hbox{{\bf/}}}\kern +.1 em}
\newcommand{\Div}{\operatorname{div}}
\newcommand{\dist}{\operatorname{dist}}
\newcommand{\Rfour}{\bar R}

\newcommand{\tthreeg}{\tilde\threeg}
\newcommand{\tcalD}{\widetilde\calD}
\newcommand{\tnabla}{\tilde \nabla}

\newcommand{\tD}{\tilde D}

\newcommand{\gammab}{\bar\gamma}
\newcommand{\chib}{\bar\chi}
\newcommand{\gb}{\bar g}
\newcommand{\Nb}{\bar N}
\newcommand{\Hb}{\bar H}
\newcommand{\Ab}{\bar A}
\newcommand{\Bb}{\bar B}
\newcommand{\betah}{{\hat\beta}}
\newcommand{\chit}{\tilde\chi}
\newcommand{\Ht}{\tilde H}
\newcommand{\Ric}{\operatorname{Ric}}
\newcommand{\supp}{\operatorname{supp}}
\newcommand{\bA}{\mathbf{A}}
\newcommand{\st}{\colon\>}

\newcommand{\hbound}{\mathring{\mathsf{H}}}
\newcommand{\cbord}{{\mathsf{C}}}
\newcommand{\pM}{\partial M}
\newcommand{\hM}{\widehat{M}}
\newcommand{\chyp}{\mathcal C}
\newcommand{\hhyp}{\mathring{\mathcal H}}
\newcommand{\bnabla}{\overline{\nabla}}
\newcommand{\maclK}{{\mathcal K}}
\newcommand{\maclKzo}{{\mathcal K}^{\bot_g}_0}
\newcommand{\maclKz}{{\mathcal K}_0}
\newcommand{\hbord}{\hbound}%
\newcommand{\cKi}{{\mycal K}_{i0}}
\newcommand{\mcO}{{\mycal O}}
\newcommand{\mcT}{{\mycal T}}
\newcommand{\mcU}{{\mycal U}}
\newcommand{\mcV}{{\mycal V}}
\newcommand{\eug}{{\frak G}}

\newcommand{\rd}{\,{ d}} 

\newcommand{\ourU}{\mathbb U}
\newcommand{\ourW}{\mathbb W}
\newcommand{\hyp}{{\mycal S}}

\newcommand{\bhyp}{\overline{\hyp}}

\newcommand{\bg}{{\overline{g}_\Sigma}}

\newcommand{\Bgamma}{{B}} 
\newcommand{\bmetric}{{b}} 

\newcommand{\Kp}{{\mathfrak g}} 
\newcommand{\KA}{p} 
\newcommand{\gthreeup}{\,{}^3g} 
\newcommand{\cO}{{\mathscr O}}
\newcommand{\arcsh}{{\rm argsh}}

\newcommand{\threeg}{\gamma}

\newcommand{\detthreeg}{{\det(\threeg_{ij})}} 
\newcommand{\lapse}{{\nu}} 
\newcommand{\shift}{{\beta}} 
\newcommand{\threeP}{{\pi}} 
\newcommand{\tildetg}{{\tilde \threeg}} 
\newcommand{\hst}{{\breve{h}}} 

\newcommand{\zn} {\mathring{\nabla}} 
\newcommand{\znabla} {\mathring{\nabla}} 
\newcommand{\mn} {M} 

\newcommand{\eg}{e.g.,}

\newcommand{\can}{{\mathrm \scriptsize can}}

\newcommand{\mcM}{{\mycal M}}
\newcommand{\mcR}{{\mycal R}}
\newcommand{\mcH}{{\mycal H}}
\newcommand{\mcK}{{\mycal K}}
\newcommand{\notreP}{{\widehat P}}
\newcommand{\ninfty}{N_\infty}
\newcommand{\bea}{\begin{eqnarray}}
\newcommand{\beaa}{\begin{eqnarray*}}
\newcommand{\bean}{\begin{eqnarray}\nonumber}
\newcommand{\kidap}{KID-hole}
\newcommand{\kidhor}{KID-horizon}
\newcommand{\kidaps}{KID-holes}
\newcommand{\bel}[1]{\begin{equation}\label{#1}}
\newcommand{\beal}[1]{\begin{eqnarray}\label{#1}}
\newcommand{\beadl}[1]{\begin{deqarr}\label{#1}}
\newcommand{\eeadl}[1]{\arrlabel{#1}\end{deqarr}}
\newcommand{\eeal}[1]{\label{#1}\end{eqnarray}}
\newcommand{\eead}[1]{\end{deqarr}}
\newcommand{\eea}{\end{eqnarray}}
\newcommand{\eeaa}{\end{eqnarray*}}
\newcommand{\nn}{\nonumber}
\newcommand{\Hess}{\mathrm{Hess}\,}
\newcommand{\Ricc}{\mathrm{Ric}\,}
\newcommand{\Riccg}{\Ricc(g)}
\newcommand{\Lpsi}{L^{2}_{\psi}}
\newcommand{\Lpsione}{\zH1_{\phi,\psi}}
\newcommand{\Lpsitwo}{\zH^{2}_{\phi,\psi}}

\newcommand{\Lpsig}{L^{2}_{\psi}(g)}
\newcommand{\Lpsioneg}{\zH1_{\phi,\psi}(g)}
\newcommand{\Lpsitwog}{\zH^{2}_{\phi,\psi}(g)}
\newcommand{\Lpsikg}[2]{\zH^{#1}_{\phi,\psi}(#2)}

\newcommand{\zg}{\mathring{g}}

\newcommand{\be}{\begin{equation}}
\newcommand{\ee}{\end{equation}}
\newcommand{\divr }{\mbox{\rm div}\,}
\newcommand{\tr}{\mbox{\rm tr}\,}
\newcommand{\ext}{{\mathrm{ext}}}
\newcommand{\J}{\delta J}
\newcommand{\source}{\delta \rho}
\newcommand{\eq}[1]{\eqref{#1}}
\newcommand{\Eq}[1]{Equation~(\ref{#1})}
\newcommand{\Eqsone}[1]{Equations~(\ref{#1})}
\newcommand{\Eqs}[2]{Equations~(\ref{#1})-\eq{#2}}
\newcommand{\Sect}[1]{Section~\ref{#1}}
\newcommand{\Lem}[1]{Lemma~\ref{#1}}

\newcommand{\zmcH }{\,\,\,\,\mathring{\!\!\!\!\mycal H}{}}

\newtheorem{defi}{\sc Coco\rm}[section]
\newtheorem{theorem}[defi]{\sc Theorem\rm}
\newtheorem{Theorem}[defi]{\sc Theorem\rm}
\newtheorem{Disc}[defi]{\sc Discussion\rm}
\newtheorem{Conjecture}[defi]{\sc Conjecture\rm}
\newtheorem{Stat}[defi]{}
\newtheorem{lemma}[defi]{Lemma}
\newtheorem{prop}[defi]{\sc Proposition\!\rm}
\newtheorem{Definition}[defi]{\sc Definition\rm}
\newtheorem{Def}[defi]{\sc Definition\rm}
\newtheorem{theor}[defi]{\sc Theorem\rm}
\newtheorem{Proposition}[defi]{\sc Proposition\rm}
\newtheorem{proposition}[defi]{\sc Proposition\rm}
\newtheorem{lem}[defi]{\sc Lemma\!\rm}
\newtheorem{Lems}[defi]{\sc Lemma\!\rm}
\newtheorem{Lemma}[defi]{\sc Lemma\!\rm}
\newtheorem{cor}[defi]{\sc Corollary\!\rm}
\newtheorem{Corollary}[defi]{\sc Corollary\!\rm}
\newtheorem{Example}[defi]{{\sc Example}\rm}
\newtheorem{Remark}[defi]{{\sc Remark}\rm}
\newtheorem{Remarks}[defi]{{\sc Remarks}\rm}
\newtheorem{examp}[defi]{{\sc Example}\rm}



\def \Reel{\mathbb{R}}
\def \C{\mathbb{C}}
\def \R {\Reel}
\def \Hyp{\mathbb{H}}
\newcommand{\mcL}{{\mycal L}}
\def \Nat{\mathbb{N}}
\def \Z{\mathbb{Z}}
\def \N {\Nat}
\def \Sphere{\mathbb{S}}

\newcommand{\bM}{\,\overline{\!M}}

\newcommand{\zHkpp}{\zHk_{\phi,\psi}}
\newcommand{\Hkpp}{H^k_{\phi,\psi}}
\newcommand{\zHk}{\zH^k}
\newcommand{\zH}{\mathring{H}}


\newcounter{mnotecount}[section]

\renewcommand{\themnotecount}{\thesection.\arabic{mnotecount}}

\newcommand{\ednote}[1]{}

\newcommand{\mcmg}{$(\mcM,\fourg)$}


\newcommand{\cem}[1]{\textcolor{bluem}{\emph{ #1}}}
\newcommand{\cbf}[1]{\textcolor{bluem}{\bf #1}}
\newcommand{\rbf}[1]{\textcolor{red}{\bf #1}}

\definecolor{bluem}{rgb}{0,0,0.5}

\definecolor{mycolor}{cmyk}{0.5,0.1,0.5,0}
\definecolor{michel}{rgb}{0.5,0.9,0.9}

\definecolor{turquoise}{rgb}{0.25,0.8,0.7}
\definecolor{bluem}{rgb}{0,0,0.5}

\definecolor{MDB}{rgb}{0,0.08,0.45}
\definecolor{MyDarkBlue}{rgb}{0,0.08,0.45}

\definecolor{MLM}{cmyk}{0.1,0.8,0,0.1}
\definecolor{MyLightMagenta}{cmyk}{0.1,0.8,0,0.1}

\definecolor{HP}{rgb}{1,0.09,0.58}

\newcommand{\opp}[1]{}

\newcommand{\jc}[1]{\mnote{{\bf jc:} #1}}
\newcommand{\ptc}[1]{\mnote{{\bf ptc:} #1}}
\newcommand{\newchange}[1]{\mnote{{\bf ptc, new change after 29 III 08:} #1}}

\newcommand{\LA} {\bf LA:}
\newcommand{\ptca}[1]{{\color{red}\mnote{{\color{black}{\bf ptc, 8.VIII.07:}
#1} }}}
\newcommand{\ptcb}[1]{{\color{red}\mnote{{\color{red}{\bf ptc, 9.VIII.07:}
#1} }}}
\newcommand{\ptcc}[1]{{\color{red}\mnote{{\color{red}{\bf ptc, 10.VIII.07:}
#1} }}}
\newcommand{\ptcd}[1]{{\color{red}\mnote{{\color{red}{\bf ptc, after 4XI07:}
#1} }}}
\newcommand{\ptce}[1]{{\color{red}\mnote{{\color{red}{\bf ptc, after 4XI07:}
#1} }}}

\newcommand{\changea}[1]{{\color{MLM}\mnote{\color{black}{\bf ptc:}
changed on 8.VIII.07}}{\color{black}{#1}}\color{black}\ }
\newcommand{\changeb}[1]{{\color{MLM}\mnote{\color{MLM}{\bf ptc:}
changed}}{\color{MLM}{#1}}\color{black}\ }
\newcommand{\changec}[1]{{\color{red}\mnote{\color{MLM}{\bf ptc:}
changed on 10.VIII.07}}{\color{MLM}{#1}}\color{black}\ }

\newcommand{\changed}[1]{{ \mnote{ {\bf ptc:}
changed}}{\sc {#1}} \ }
\newcommand{\added}[1]{{ \mnote{ {\bf ptc:} added
}}{\sc{#1}} \ }

\newcommand{\adda}[1]{{\color{MDB}\mnote{\color{MDB}{\bf ptc:} added
who knows when} {\color{MDB}{#1}}}\color{black}\ }
\newcommand{\addc}[1]{{\color{MDB}\mnote{\color{MDB}{\bf ptc:} added
on 10.VIII.07}}{\color{MDB}{#1}}\color{black}\ }
\newcommand{\addd}[1]{{\color{MDB}\mnote{\color{MDB}{\bf ptc:} added
after 4XI07}}{\color{MDB}{#1}}\color{black}\ }

\newcommand{\reworda}[1]{{\color{green}\mnote{\color{green}{\bf ptc:}
reworded on 8.VIII.07}}{\color{green}{#1}}\color{black}\ }

\newcommand{\mcMX}{\mcM_X}

\newcommand{\pdoc}{\partial \doc}
\newcommand{\loc}{{\textrm{loc}}}

\newcommand{\xflat}{X^\flat}
\newcommand{\kflat}{K^\flat}
\newcommand{\yflat}{Y^\flat}
\newcommand{\Sext}{\hyp_{\mathrm{ext}}}
\newcommand{\Mext}{\mcM_{\mathrm{ext}}}
\newcommand{\Mextl}{\Mext^\lambda}
\newcommand{\Mextls}{\Mext^{\lambda_*}}
\newcommand{\doc}{\langle\langle \mcMext\rangle\rangle}
\newcommand{\docl}{\langle\langle\mcMext^\lambda\rangle\rangle}
\newcommand{\docls}{\langle\langle \mcMext^{\lambda_*}\rangle\rangle}
\newcommand{\doclo}{\langle\langle \mcMext^{\lambda_1}\rangle\rangle}
\newcommand{\doclt}{\langle\langle \mcMext^{\lambda_2}\rangle\rangle}
\newcommand{\Tau}{\tau}

\newcommand{\bcM}{{\,\,\,\overline{\!\!\!\mcM}}}
\newcommand{\tcM}{\,\,\,\,\widetilde{\!\!\!\!\cM}}
\newcommand{\hJ}{{\hat J}}
\newcommand{\sthd}{{}^{\star_g}}
\newcommand{\hG}{\widehat \Gamma}
\newcommand{\hD}{\widehat D}
\newcommand{\hxi}{\hat \xi}
\newcommand{\hxib}{\hat{\xib}}
\newcommand{\heta}{\hat \eta}
\newcommand{\hetab}{\hat{\etab}}
\newcommand{\home}{\hat{\omega}}
\newcommand{\homb}{\hat{\underline\omega}}

\newcommand{\cimkd}{{\mycal C}^\flat}
\newcommand{\cimku}{{\mycal C}^\sharp}
\newcommand{\proj}{\textrm{pr}}
\newcommand{\Mclosed}{\bcM}
\newcommand{\cg}{\,{\tilde {\!g}}}

\newcommand{\cM}{\mycal M}
\newcommand{\cN}{\mycal N}
\newcommand{\cE}{\mycal E}

\newcommand{\cAp}{{\mycal A}_{\mbox{\scriptsize phg}}}
\newcommand{\cApM}{\cAp(M)}
\newcommand{\stsg}{{\mathfrak g}}
\newcommand{\tf}{\widetilde f}
\newcommand{\complementaire}{\complement}
\newcommand{\remarks}{{\bf Remarks : }}

\newcommand{\mcB}{{\mycal B}}
\newcommand{\decal}{{\mycal D}}
\newcommand{\cC}{{\mycal C}}
\newcommand{\cB}{{\mycal B}}
\newcommand{\cG}{{\mycal G}}
\newcommand{\cCak}{{\cC}^\alpha_k}
\newcommand{\cU}{{\mycal  U}}
\newcommand{\backmg}{b} 
\newcommand{\Id}{\mbox{\rm Id}} 
\newcommand{\sconst}{\mbox{\scriptsize\rm const}} 

\newcommand{\chnindex}[1]{\index{#1}}
\newcommand{\chindex}[1]{\index{#1}}
\newcommand{\bhindex}[1]{\chindex{black holes!#1}}

\newcommand{\tA}{\theta_\Al}
\newcommand{\Hau}{{\mathfrak H}}
\newcommand{\Ar}{\mbox{\rm Area}}
\newcommand{\Arm}{\mbox{\rm Area}_{S_2}}
\newcommand{\Al}{{\mathcal Al}}
\newcommand{\Hausone}{\Hau1}
\newcommand{\Htwohone}{\Hau^{2}}
\newcommand{\calS}{{\mathscr S}}
\newcommand{\BA}{B_\Al}
\newcommand{\Leb}{{\mathfrak L}}

\newcommand{\eqs}[2]{\eq{#1}-\eq{#2}}

\newcommand{\cW}{{\mycal W}}

\newcommand{\cA}{{\mycal A}}

\def\nen{\nonumber}
\def\Tau {{\mycal  T}}

\newcommand{\pihyp}{\partial_\infty\hyp}
\newcommand{\cMext}{{\mcM}_{\ext}}
\newcommand{\timkd}{{\mycal T}^\flat}
\newcommand{\timku}{{\mycal T}^\sharp}

\newcommand{\ro}{\rho}
\newcommand{\roo}{\bar{\ro}}

\newcommand{\les}{\lesssim}
\newcommand{\ges}{\gtrsim}
\def\piT{{\,^{(\T)}\pi}}
\def\Qr{\mbox{Qr}}
\def\flux{{\mbox{Flux}}}
\def\pa{\partial}
\def\onab{\overline{\nabla}}
\def\Db{{\bf {\bar {\,D}}}}
\def\Dh{{\bf {\hat {\,{D}}}}}
\def\laph{\hat{\Delta}}
\def\Null{\dot{\NN}^{-}}
\def\EE{{\mycal  E}}
\def\GG{{\mycal  G}}
\def\OO{{\mycal  O}}
\def\QQ{{\mycal  Q}}
\def\UU{{\mycal  U}}
\def\VV{{\mycal  V}}
\def\TT{{\mycal  T}}
\def\exp{\,\mbox{exp}}
\def\P{{\bf P}}
\def\vphi{\varphi}

\def \endprf{\hfill  {\vrule height6pt width6pt depth0pt}\medskip}
\def\emph#1{{\it #1}}
\def\textbf#1{{\bf #1}}
\newcommand{\Us}{\mbox{$U\mkern-13mu /$\,}}
\newcommand{\eql}{\eqlabel}
\def\medn{\medskip\noindent}

\def\Db{{^{(\Lb)}  D}}

\def\a{{\alpha}}
\def\b{{\beta}}
\def\ga{\gamma}
\def\Ga{\Gamma}
\def\eps{\epsilon}
\def\La{\Lambda}
\def\si{\sigma}
\def\Si{\Sigma}
\def\ro{\rho}
\def\th{\theta}
\def\ze{\zeta}
\def\nab{\nabla}
\def\varep{\varepsilon}

\def\aa{{\underline{\a}}}
\def\bb{{\underline{\b}}}
\def\bb{{\underline{\b}}}
\def\omb{{\underline{\om}}}
\def\Lb{{\underline{L}}}
\def\aaa{{\bold a}}

\newcommand{\trchb}{\tr \chib}
\newcommand{\chih}{\hat{\chi}}
\newcommand{\xib}{\underline{\xi}}
\newcommand{\etab}{\underline{\eta}}
\newcommand{\chibh}{\underline{\hat{\chi}}\,}
\def\chih{\hat{\chi}}
\def\trch{\mbox{tr}\chi}
\def\tr{\mbox{tr}}
\def\Tr{\mbox{Tr}}
\def\Xb{{\underline X}}
\def\Yb{{\underline Y}}

\def\rhoc{\check{\rho}}
\def\sic{\check{\si}}
\def\bboc{\check{\bb}}
\def\muc{\tilde{\mu}}

\def\AA{{\mycal  A}}
\def\BB{{\mycal  B}}
\def\MM{{\mycal  M}}
\def\NN{{\mycal  N}}
\def\II{{\mycal  I}}
\def\FF{{\mycal  F}}
\def\HH{{\mycal  H}}
\def\JJ{{\mycal  J}}
\def\KK{{\mycal  K}}
\def\Lie{{\mycal  L}}
\def\DD{{\mycal  D}}
\def\PP{{\mycal  P}}
\def\HH{{\mycal  H}}

\def\LL{{\mathcal L}}

\def\A{{\bf A}}
\def\B{{\bf B}}
\def\F{{\bf F}}
\def\I{{\bf I}}
\def\J{{\bf J}}
\def\M{{\bf M}}
\def\L{{\bf L}}
\def\O{{\bf O}}
\def\Q{{\bf Q}}
\def\R{{\mathbb R}}
\def\U{{\bf U}}
\def\S{{\bf S}}
\def\K{{\bf K}}
\def\g{{\bf g}}
\def\t{{\bf t}}
\def\u{{\bf u}}
\def\p{{\bf p}}
\def\LLb{{\bf \Lb}}

\def\Ab{\underline{A}}
\def\Eb{\underline{E}}
\def\Bb{\underline{B}}
\def\Hb{\underline{H}}
\def\Rb{\underline{R}}
\def\Vb{\underline{V}}
\def\AAA{{\Bbb A}}
\def\RRR{{\Bbb R}}
\def\MMM{{\Bbb M}}
\def\T{{\Bbb T}}
\def\To{\T}

\def\Aoc{\check{A}}
\def\Roc{\check{R}}
\def\nabb{\overline{\nab}}
\newcommand{\nabbb}{\mbox{$\nabla \mkern-13mu /$\,}}
\def\Ml{{\mycal  M}_{\ell}}
\def\intws{{\,\,\,\int_0^s \mkern-30mu \om}\,\,\,\,}
\def\expp{\mbox{exp}}

\def\B{{\bf B}}

\newcommand{\changedX}{K}

\newcommand{\hahyp}{\,\,\widehat{\!\!\hyp}}%

\def\lap{\Delta}
\def\pr{\partial}
\newcommand{\ddd}{\nab}

\def\c{\cdot}

\def\Da{{^{(A)}\hskip-.15 pc \D}}

\newcommand{\distb}{d_\backmg}
\newcommand{\hmcK}{\widehat \mcK}
\newcommand{\ptcx}[1]{\ptc{#1}}

\def\hot{\widehat{\otimes}}
\def\lot{\mbox{l.o.t.}}

\renewcommand{\div}{\mbox{div }}
\newcommand{\divv}{\mbox{div} }
\def\err{\mbox{Err}}
\newcommand{\lapp}{\mbox{$\bigtriangleup  \mkern-13mu / \,$}}

\newcommand{\piX}{\,^{(X)}\pi}
\def\diag{{\mbox{diag}}}
\def\Flux{\mbox{Flux}}
\def\En{\mbox{En}}
\def\ub{{\underline u}}
\def\dual{{\,^\star \mkern-3mu}}
\def\2{{\overline 2}}
\def\NI{\noindent}
\def\Cb{\underline C}

\newcommand{\beqa}{\begin{eqnarray}}
\newcommand{\eeqa}{\end{eqnarray}}

{\catcode `\@=11 \global\let\AddToReset=\@addtoreset}
\AddToReset{figure}{section}
\renewcommand{\thefigure}{\thesection.\arabic{figure}}

\title{On higher dimensional black holes with abelian isometry group}

\begin {abstract}
We consider $(n+1)$--dimensional,  stationary, asymptotically
flat, or Kaluza-Klein asymptotically flat  black holes,  with
an abelian $s$--dimensional subgroup of the isometry group
satisfying an orthogonal integrability condition. Under
suitable regularity conditions we prove that the area of the
group orbits is positive on the domain of outer communications
$\doc$, vanishing only on the boundary $\pdoc$ and on the
``symmetry axis" $\mcA$. We further show that the orbits of the
connected component of the isometry group are timelike
throughout the domain of outer communications. Those results
provide a starting point for the classification of such black
holes. Finally, we show non-existence of zeros of static
Killing vectors on degenerate Killing horizons, as needed for
the generalisation of the static no-hair theorem to higher
dimensions.
\end {abstract}
\maketitle

\tableofcontents

\mainmatter

\section{Introduction}
 \label{Sintro}

  \ptc{watch out with condition 5.13 in Bourguignon, in the proof of Thm 5.1, and condition on s in Thm 6.2 and lemma 6.3 }

In this work we study the global structure of stationary
 space-times with $s+1$, $s\ge 0$,  commuting Killing vectors
$K_\kl \mu$, $\mu=0,\ldots s$, satisfying  the \emph{orthogonal integrability} condition:
\begin{equation}
\label{intcond2}
 \forall\  \mu=0,\ldots, s \qquad dK_\kl{\mu }\wedge K_\kl{ 0}
 \wedge \ldots \wedge
K_\kl{{s}} =0 \;.
\end{equation}
This class  includes the Kerr metrics,  the ``black strings"
Kerr$\times S^1$ and other abelian Kaluza-Klein black-holes as
in~\cite{BMG},   the Emparan-Reall ``black
rings"~\cite{EmparanReall},  a subset of the Myers-Perry black
holes~\cite{myersperry}, as well as the Elvang-Figueras ``black
Saturns"~\cite{EF}.

Condition~\eq{intcond2} automatically holds in, say vacuum,
$(n+1)$--dimensional space-times when $s=n-2$ and when the
``axis" $\mcA$ defined below is non-empty~\cite{CarterJMP}
(compare~\cite{ERWeyl,ChCo}). However, one might wish to
consider metrics where \eq{intcond2} is imposed as a
restrictive condition, without necessary assuming that $s=n-2$.

A prerequisite  to the classification of the above
geometries~\cite{CarterlesHouches,BMG} (see also~\cite{ChCo})
is the understanding of the global structure of the domain
of outer communications%
\footnote{See Section~\ref{SKK} and~\cite{ChCo}
for terminology and definitions.\label{f1}}
$\doc$: one needs a product structure of $\doc$ with respect to the
action of the stationary Killing vector field, information
about $\pi_1(\doc)$,  knowledge of the causal character of the
orbits of the Killing vectors. The aim of this work is to
settle some of those issues.

Specifically, one of the key issues is
the analysis of the zero level set of the ``area function" $W$,
defined as
$$
 W:= -\det \left(\fourg(K_\kl \mu, K_\kl \nu)_{\mu,\nu=0,\ldots ,s}\right)
 \;.
 $$
Indeed the uniqueness theory of such black
holes~\cite{BMG,CarterlesHouches} uses $\sqrt W$ as one of the
coordinates on the quotient of the domain of outer
communications by the isometry group. Clearly a function $W$
changing sign would invalidate the whole approach. Our first
main result, Theorem~\ref{Tdoc2} below, asserts that, under suitable
regularity and asymptotic conditions,  the area function $W$
vanishes within the domain of
outer communications
\emph{only} on the \emph{axis}
\bel{Adef}
 \mcA:= \{p\in \mcM \ | \ Z(p)=0\}
  \;,
\ee
where
$$
 Z:= \det \left(\fourg(K_\kl i, K_\kl j)_{i,j=1,\ldots ,s}\right)
 \;.
 $$
The proof relies heavily on the analysis in~\cite{ChCo}, as
well as on the results in~\cite{CGS} which are reviewed in our
context in Section~\ref{sghsc}.

Next, inspection of the uniqueness arguments
in~\cite{Weinstein1,HY,IdaMorisawa,RobinsonKerr} shows that
serious difficulties arise there if the orbits of the isometry
group cease to be timelike on $ \mcA$.  The second main result
of our work is Theorem~\ref{TdocA} below, that this does not
occur.

The simplest non-trivial abelian isometry Lie group is $\R$,
then $s=0$ and the orthogonal integrability condition
\eq{intcond2} is known as the staticity condition. Now, the
generalization of the uniqueness theory of static
asymptotically flat black holes requires the non-vanishing of
the static Killing vector on degenerate components of the event
horizon. We prove this in Proposition~\ref{Ponce2x}; this is
the third main result in this paper.

For reasons discussed in detail shortly, we work in the
framework of manifolds which are asymptotically flat in a
Kaluza-Klein sense, as defined below; manifolds which are
asymptotically flat in the usual sense occur as a special case
of our analysis.

\section{Kaluza-Klein asymptotic flatness}
 \label{SKK}

Consider an  $(n+1)$--dimensional space-time $(\mcM,\fourg)$
which is asymptotically flat in the usual sense, as e.g.
in~\cite{ChMaerten}. It follows from the analysis there that
there exists a homomorphism from the connected component $G_0$
of the identity of the group of isometries of $(\mcM,\fourg)$
to a subgroup of the Lorentz group, constructed using the
leading order behavior of the Killing vectors of
$(\mcM,\fourg)$. Assuming that the ADM four-momentum of
$(\mcM,\fourg)$ is timelike, arguments similar to those leading
to \eq{rcalc} below show that the dimension of any commutative
subgroup of $G_0$ does not exceed $n/2+1$, where $n/2$ arises
from the rank of $SO(n)$,  while ``$+1$" comes from a possible
time-translation. This implies that the hypothesis of
asymptotic flatness is compatible with the condition $s=n-2$
only in space-dimension $n$ equal to three and four.

However, in the context of Kaluza-Klein theories, there are
situations of interest which are not asymptotically flat and to
which the current analysis applies. A trivial example is given
by space-times of the form $(\mcM\times S^1, \fourg^{\kl
{n+1}}+dx^2)$, where $\fourg^{\kl {n+1}}$ is an
$(n+1)$--dimensional asymptotically flat, say Ricci flat,
metric (e.g. Kerr, or Myers-Perry, or Emparan-Reall). In this
trivial product case the Einstein equations reduce to the ones
for the quotient metric  $(\mcM , \fourg^{\kl {n+1}})$, so
there is no point in generalising. Now, one can imagine
situations where the higher-dimensional metric asymptotes to a
product solution, but does \emph{not} lead to a metric
satisfying the required hypotheses after passing to the
quotient. For example, the quotient metric associated with a
vacuum metric will not satisfy the positive energy condition in
general. So there appears to be some interest to relax the
asymptotic flatness condition,  perhaps to show eventually that
the resulting solutions must be trivial products.

With this motivation in mind, we shall say that $\Sext$ is a
\emph{Kaluza-Klein asymptotic end}, or \emph{asymptotic end}
for short, if $\Sext$ is diffeomorphic to $\left(\R^n\setminus
\overline B(R)\right) \times N$, where $\overline B(R)$ is a
closed coordinate ball of radius $R$, and $N$ is a compact
manifold.
Let $\mathring k$ be a fixed Riemannian metric on $N$, and let
$\mathring g = \delta \oplus \mathring k$, where $\delta $ is
the Euclidean metric on $\R^n$.

We shall say that a Riemannian metric $g$ on $\Sext$ is
 \emph{Kaluza-Klein asymptotically flat}, or
 \emph{$KK$--asymptotically flat} for short, if there exists
 $\alpha>0$ and $k\ge 1$ such that for $0\le \ell \le k$ we
\bel{meddec} \zD_{i_1}\ldots \zD_{i_\ell} (g_{jk}-\mathring
g_{jk}) = O(r^{-\alpha-\ell})
 \;,
\ee
where $\zD$ denotes the Levi-Civita connection of $\zg$, and
$r$ is the radius function in $\R^n$, $r:=\sqrt{(x^1)^2+\ldots
(x^n)^2}$, with the $x^i$'s being any Euclidean coordinates of
$(\R^n,\delta)$.   We shall say that a general relativistic
initial data set $(\Sext,g,K) $   is \emph{Kaluza-Klein
asymptotically flat}, or \emph{$KK$--asymptotically flat}, if
$(\Sext,g)$ is $KK$--asymptotically flat and if for  $0\le \ell
\le k-1$ we have
\bel{Kdec} \zD_{i_1}\ldots \zD_{i_\ell} K_{jk} =
O(r^{-\alpha-1-\ell})
 \;.
\ee

The above reduces to the usual notion of asymptotic flatness
when $N$ is a set containing one point. So an asymptotically
flat initial data set is also $KK$--asymptotically flat.

Consider a space-time $\mcM$ containing a $KK$--asymptotically
flat end $(\Sext,K,g)$, and suppose that there exists on $\mcM$
a Killing vector field $X$ with complete orbits. Then $X$ will
be called
 \emph{stationary} if $X$ approaches the timelike unit normal to $\Sext$ when
 one recedes to infinity along $\Sext$.
 $(\mcM,\fourg)$ will then be called stationary. Such a
 space-time will then be called $KK$--asymptotically flat.

Similarly to the standard asymptotically flat case, we set
$$
 \Mext:= \cup_{t\in \R} \phi_t[X](\Sext)\;,
$$
where  $\phi_t[X]$ denotes the flow of $X$. Assuming
stationarity, the \emph{domain of outer communications} is
defined as in~\cite{ChWald1,ChCo}:
$$
 \doc:= I^-(\Mext)\cap I^+(\Mext)
 \;.
$$

\section{Simple connectedness of the orbit manifold}
 \label{sghsc}

In the current context, and in  higher dimensions $n\ge 4$,
simple connectedness holds for asymptotically flat  globally
hyperbolic domains of outer communications satisfying the null
energy condition
\bel{NEC}
 R_{\mu\nu}Y^\mu Y^\nu\geq 0 \,\,\text{ for null }\,\, Y^\mu
  \;.
\ee
Indeed, the analysis
in~\cite{ChWald,galloway-topology,Galloway:fitopology,FriedmanSchleichWitt},
carried-out there in dimension $3+1$, is independent of
dimensions. However, as already discussed, asymptotic flatness
imposes $n=3$ or $4$ if one wishes to derive, rather than
impose, the orthogonal integrability condition \eq{intcond2}.
In any case, KK-asymptotically flat solutions will not be
simply connected in general, as demonstrated by the
Schwarzschild$\times \T^m$ ``black branes".

Now, whenever simple connectedness fails, the twist potentials
might fail to exist  and the whole reduction
process~\cite{BMG}, that relies on their existence, breaks
down. It turns out  that the quotient space $\doc/\T^s$ remains
simply connected for $KK$--asymptotically flat models, which
justifies existence of twist potentials whenever $\doc/\T^s$ is
a manifold. Moreover, simply connectedness of $\doc/\T^s$ will
be used below to show that the area function has no zeros on
$\doc$.

Indeed, a variation upon the usual topological censorship
arguments~\cite{Galloway:fitopology,ChWald,FriedmanSchleichWitt}
gives:

\begin{Theorem}[\cite{CGS}]
 \label{TCdsc} Let $(\mcM,\fourg)$ be a space-time satisfying the null energy
condition, and containing a $KK$--asymptotically flat end
$\Sext$. Suppose that $\doc$ is globally hyperbolic, and that
there exists an action of $G=\R\times G_s$
 on $\doc$ by isometries which, on $\Mext\approx \R\times
 \Sext$, takes the form
$$
\R\times G_s \ni (\tau,g):\quad  (t,p)\mapsto (t+\tau,g\cdot p)
 \;.
$$
We assume moreover that the generator of the $\R$ factor of $G$
approaches the unit timelike normal to $\Sext$ as one recedes
to infinity. If $\Sext/G_s$ is simply connected, then so is
$\doc/G_s$.
\end{Theorem}
\medskip

At the heart of Theorem~\ref{TCdsc} lies
Proposition~\ref{Pnotrapped} below.  Before describing the
result, some definitions are in order. Consider a spacelike
manifold $S\subset \mcM$ of co-dimension two, and assume that
there exists a smooth unit spacelike vector field $n$ normal to
$S$ such that the vector fields $\pm n$ lie in distinct
components of the bundle of spacelike vectors normal to $S$; we
shall call \emph{outwards} the  component met by  $n$, and the
other one \emph{inwards}.    At every point $p\in S$ there
exists then a unique future directed null vector field $n^+$
normal to $S$ such that $\fourg(n,n^+)=1$, which we shall call
the \emph{outwards future null normal} to $S$. The
\emph{inwards future null normal} $n^-$ is defined   by the
requirement that $n^-$ is null, future directed, with
$\fourg(n,n^-)=-1$. In an asymptotically flat, or
$KK$--asymptotically flat region the inwards direction at
$\{r=R\}$ is defined to by $dr(n)<0$.

We define the \emph{null future inwards and outwards mean
curvatures $\theta^\pm$ of $S$}   as
\bel{thetdef}
 \theta^\pm :=\tr_\gamma( \nabla n^\pm)
 \;,
\ee
where $\gamma$ is the metric induced on $S$. In \eq{thetdef}
the symbol $n^\pm$ should be understood as representing any
extension of the null normals $n^\pm$ to a neighborhood of $S$,
and the definition is independent of the extension chosen.

We shall say that $S$ is \emph{weakly outer future trapped} if
$\theta^+\le 0$.  The notion of \emph{weakly inner future
trapped} is defined by requiring $\theta^-\le 0$.  A similar
notion of \emph{weakly outer or inner past trapped} is defined
by considering the divergence of past pointing null normals. We
will say \emph{outer future trapped} if $\theta^+< 0$, etc.

Let $t$ be a time function on $\mcM$, and let $\gamma:[a,b]\to
\mcM$ be a causal curve. The \emph{time of flight $t_\gamma$ of
$\gamma$} is defined as
$$
 t_\gamma= t(\gamma(b))-t(\gamma(a))
 \;.
$$

In what follows we will  need the following, also proved
in~\cite{CGS}:

\begin{Proposition}
[\protect\cite{CGS}, Proposition~5.3] \label{Pnotrapped} Let
$(\mcM,\fourg)$ be a stationary, asymptotically flat, or
$KK$--asymptotically flat globally hyperbolic space-time
satisfying the null energy condition. Let $S\subset \doc$ be
future inwards marginally trapped.  There exists a constant
$R_1$ such that for all $R_2 \ge R_1$ there are no future
directed null geodesics starting inwardly at $S$, ending
inwardly at $\{r=R_2\}\subset \Mext$, and locally minimising
the time of flight,
\end{Proposition}

\section{The structure of the domain of outer communications}
 \label{ssNzK}

We wish, here, to point out a set of hypotheses
which allows one to establish the $KK$--asymptotically flat
counterpart of the Structure Theorem
of~\cite[Section~4.2]{ChCo}, Theorem~\ref{Tgt} below. This
shows in particular that the action of the isometry group
on $\doc$ is of the form assumed in Theorem~\ref{TCdsc}.

In this section we  assume the existence of a connected
subgroup $G=\R\times G_s$ (here the subscript ``$s$" stands for
``spacelike") of the group of isometries of $(\mcM,\fourg)$,
where $G_s=G_1\times G_2$ is a compact group, with the
following action in the asymptotic region: let us write $\Mext$
as $\R\times \Sext$, where the $\R$ factor of $G$ acts by
translations on the $\R$ factor of $\R\times \Sext$. Each of
$G_s$, $G_1$ and $G_2$ is allowed to be trivial, and neither is
assumed to be commutative. Recalling that
\bel{afcp}
(\Sext,\zg)=\left(
\left(\R^n\setminus \overline B(R)\right) \times N,  \delta
\oplus \mathring k\right)
 \;,
\ee
we assume that $G_1$ is a
subgroup of $SO(n)$ acting by rotations of the flat metric
$\delta$ on $\R^n\setminus \overline B(R)$ and trivially on
$N$, and that $G_2$ acts on the $N$ factor by isometries of
$\mathring k$ and trivially on $\R^n\setminus \overline B(R)$.
Finally, we suppose that the Killing vector tangent to the $\R$
factor of $\Mext$, and denoted by  $\Kz$, is timelike on
$\Mext$. Note that all the remaining Killing vectors, denoted
by $K_\kl i$, if any, have spacelike or trivial orbits in
$\Mext$.

In the asymptotically flat case, the existence of coordinates
as in \eq{afcp} can be derived from asymptotic flatness if a timelike
ADM four-momentum of $\Sext$ is assumed~\cite{ChMaerten}. It
would be of interst to determine whether or not this remains
true in the $KK$--asymptotically flat setup.

The following definition is a direct generalisation of the one
in~\cite{ChCo}:

\begin{Definition}
 \label{Dmain} Let $(\mcM,\fourg)$ be a space-time containing a
$KK$--asymptotically flat end $\Sext$, and let  $\changedX $ be
a stationary Killing vector field  on $\mcM$.
We will say that $(\mcM,\fourg,\changedX)$ is $\mbox{\rm {\regular}}$%
\index{$\mbox{\rm {\regular}}$}
if $\changedX $ is complete, if the domain of outer communications
$\doc$ is globally hyperbolic, and if $\doc$ contains a spacelike,
connected, acausal hypersurface $\hyp\supset\Sext $,%
\index{$\hyp$}
the  closure $\ohyp $ of which is a topological manifold with boundary,
consisting of  the union of a compact set and of a finite number of
asymptotic ends, such that the boundary $ \pohyp:= \ohyp \setminus \hyp$
is a  topological manifold satisfying
\bel{subs}
\pohyp \subset \mcE^+:= \partial \doc \cap I^+(\Mext)
 \;,
 \ee
with $\pohyp$ meeting every generator of $\mcE^+$ precisely
once. See~Figure~\ref{fregu}.
\begin{figure}[ht]
\begin{center} { \psfrag{Mext}{$\phantom{x,}\Mext$}
\psfrag{H}{ } \psfrag{B}{ }
\psfrag{H}{ }
 \psfrag{pSigma}{$\!\!\pohyp\qquad\phantom{xxxxxx}$}
\psfrag{Sigma}{ $\hyp$ }
 \psfrag{toto}{$\!\!\!\!\!\!\!\!\!\!\doc$}
 \psfrag{S}{}
\psfrag{H'}{ } \psfrag{W}{$\mathcal{W}$}
\psfrag{scriplus} {} 
\psfrag{scriminus} {} 
 \psfrag{i0}{}
\psfrag{i-}{ } \psfrag{i+}{}
 \psfrag{E+}{ $\phantom{.}{\mycal E}^+$}
{\includegraphics{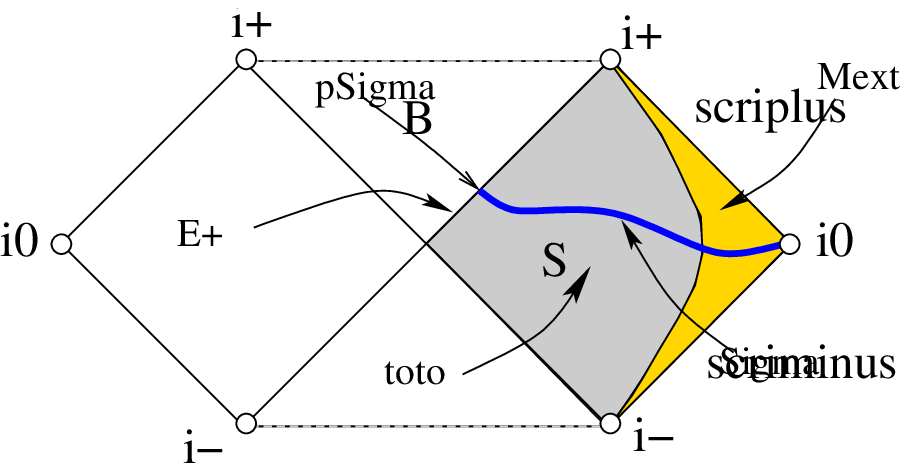}}
}
\caption{The hypersurface $\hyp$ from the definition of \regular ity.
\protect\label{fregu}}
\end{center}
\end{figure}
\end{Definition}

\newcommand{\Sts}{\hat S_{\tau,\sigma}}%
\newcommand{\hatt}{\hat t}%
The proof of the Structure
Theorem~\cite[Theorem~4.5]{ChCo} carries over with only
trivial modifications to the current setting:

\begin{Theorem}[Structure theorem]%
\index{structure theorem}%
 \label{Tgt}
Suppose that $(\mcM,\fourg)$ is {\aregular}  space-time
invariant under an action of $G=\R\times G_s$ as above. There
exists on $\doc $ a smooth time function $t$, invariant under $
G_s$, which, together with the flow of the Killing vector $\Kz$
tangent to the orbits of the $\R$ factor of $G$, induces the
diffeomorphisms
\bel{gooddecomp}
 \doc \approx \R \times \hypo
 \;,\qquad
\overline{ \doc}{\cap I^+(\Mext)} \approx \R \times \ohypo
 \;,
\ee
where $\hypo:=t^{-1}(0)$ is $KK$--asymptotically flat,
(invariant under $G_s$), with the boundary $\partial \ohypo$
being a compact cross-section of $\mcE^+$.  The smooth
hypersurface with boundary $\ohypo$ is acausal, spacelike
up-to-boundary, and the flow of $\Kz $ is a translation along
the $\R$ factor in \eq{gooddecomp}.
\end{Theorem}

\medskip

\section{The area function away from the axis}
 \label{sAfaa}

In this section we prove a generalization of~\cite[Theorem~5.4
and 5.6]{ChCo}. The main issue is, that an essential ingredient
of the proof in~\cite{ChCo} is simple connectedness of $\doc$,
which is not expected for $KK$--asymptotically flat space-times
with internal space $N=\T^k$. (We emphasize that we \emph{do
not} assume this form of $N$ in this section, but this is the
model which seems to be of main interest for applications of
this work.) Here we use instead the closely related
Proposition~\ref{Pnotrapped}, obtaining:

\begin{Theorem}
 \label{Tdoc2} Under the hypotheses of Theorem~\ref{TCdsc},
suppose further that $G=\R\times \T^{s }$ with
$s+1$--dimensional principal orbits, $0\le s\le n-2$. Assume
moreover that \underline{either} $(\mcM,\fourg)$ is analytic,
\underline{or} that $s=n-2$ and $(\mcM,\fourg)$ is \regular. If
the orthogonal integrability condition \eq{intcond2} holds,
then the function
\bel{Wdef2}
 W:= -\det\Big( \fourg(K_\kl \mu,K_\kl
 \nu)\Big)_{\mu,\nu=0,\ldots,s }
\ee
is strictly positive on $\doc\setminus\mcA$, and vanishes  on $\pdoc\cup \mcA$.
\end{Theorem}

\proof
We wish to adapt the proof of~\cite[Theorems~5.4 and 5.6]{ChCo}
to the current setting. Let us show, first, that the existence
of a non-empty, closed, embedded,  null hypersurface
$S^+\subset
\doc$, invariant under $ \T^s$, is incompatible with what we
know about the topology of $\doc$:

\emph{If} $\doc/\T^s$ is a smooth manifold, and \emph{if}
$S^+/\T^s$ is a non-empty, closed,   embedded hypersurface in
$\doc/\T^s$, one can proceed as follows. Let $\gamma$ be a
closed path with strictly positive $S^+$--intersection number,
as constructed in the last step of the proof
of~\cite[Theorem~5.4]{ChCo}. Then $\pi(\gamma)$, where
$\pi:\doc\to \doc/\T^s$ is the projection map, has strictly
positive $S^+ /\T^s$--intersection number, which contradicts
simple connectedness of $\doc/\T^s$, and proves the result.

However, it is not clear that both assumptions of the previous
 paragraph will hold in general, in which case the following
 argument applies: Suppose that $S^+$
 is non-empty, let
 $\hyp\supset \Sext$ be any $KK$--asymptotically flat level set
 of a Cauchy time function in $\doc$. Choose $R_0$ large enough
 so that the spacelike manifold $S_0:=\{r=R_0\}\cap \hyp$ is
 both past and future inwards trapped, and that any causal
 curve from $S_0$ to $\partial \Mext$ takes at least a
 coordinate time one before reaching $\partial\Mext$. Let $R_2\ge
 \max(R_0.R)$, where $R$ is as in Proposition~\ref{Pnotrapped}.
 There exists a future directed causal curve  from $S_0$ to
 $\{r=R_2\}$ which starts in the inwards direction at $S_0$,
 leaves $\Mext$, meets $S^+$, and returns to $\{r=R_2\}$. This
 shows that the set
 $$
  \Omega:=\{\gamma \ | \ \mbox{$\gamma$ is a causal curve from $S_0$ to $\{r=R_2\}$ meeting $S^+$}\}
 $$
 is non-empty. Let $t_\gamma$ denote the coordinate arrival time of
 $\gamma\in \Omega$ to $\{r=R_2\}$ then $t_\gamma\ge t|_{S_0}+2$. Let
 $\gamma_i\in \Omega$ be any sequence such that
 $$
 t_{\gamma_i}\to \inf_{\gamma\in \Omega}t_\gamma \ge  t|_{S_0}+2
 \;.
 $$
 Let $\gamma_*$ be an accumulation curve of the $\gamma_i$'s,
 global hyperbolicity implies that $\gamma_*:[a,b]\to \mcM$ is
   a non-trivial null geodesic from $S_0$ to $\{r=R_2\}$
   without, in the terminology of~\cite{Galloway:fitopology},
   null $S_0$--focal points on $[a,b)$, inwards
 directed at $S_0$, and providing a local minimum of time of
 flight between $S_0$ and $\{r=R_2\}$. This, however, contradicts
 Proposition~\ref{Pnotrapped}, hence $S^+$ is empty.

In the analytic case, the arguments of the proof of
\cite[Theorem~5.4]{ChCo} show that the existence of zeros of
$W$ in $
\doc$ leads to the existence of an embedded hypersurface $S^+$
as above, contradicting what has just been said.
 \ptc{should I worry here about 5.13 in ChCo?}

If $s=n-2$ and $n=4$, the proof of \cite[Theorem~5.6]{ChCo}
applies.

In what follows we assume that the reader is familiar with the
notation, and arguments, of the proof of
\cite[Theorem~5.6]{ChCo}.

Now, if $s=n-2$ and $n\ge 5$, one needs to exclude the
possibility that the leaves $C_q$  pass through points on
$\mcA$ which are intersection points of two or more axes of
rotation. Suppose, for contradiction, that there exists such a
point.  Let $p_0\in
\doc$ be a corresponding point where the axes of rotation meet,
then there exists a null $\T^s$--invariant (not necessarily
embedded) hypersurface $\hat S_p$, totally geodesic in $\doc$,
passing through $p_0$. Choose a basis $\{K_\kl 1,\ldots,K_\kl
s\}$ of the Lie algebra of $\T^s$ such that the  Killing
vectors $\{K_\kl {r+1},\ldots,K_\kl s\}$ form a largest
linearly independent subfamily at $p_0$, while $\{K_\kl
1,\ldots,K_\kl r\}$ vanish at $p_0$. Let $\T^{s-r}$ denote the
group of isometries generated by $\{K_\kl {r+1},\ldots,K_\kl
s\}$, then $ \mcM/\T^{s-r}$ is a smooth manifold near $p_0$,
and $\hat S_p/ \T^{s-r}$ is a smooth hypersurface there. We can
equip $ \mcM/\T^{s-r}$ with the quotient space-metric: for
$Z,W\in T\mcM/\T^{s-r}$,
$$
 \gamma(Z,W) := \fourg (\hat Z,\hat W)- h^{\kl i \kl j} \fourg(\hat Z,K_\kl i)\fourg(\hat W,K_\kl j)
 \;,
$$
where $h^{\kl i \kl j}$ is the matrix inverse to the matrix
$g(K_\kl i,K_\kl j)$, with $i,j=k+1,\ldots, s$, and $(\hat
W,\hat Z)$ are any vectors in $T\mcM$ which project on $(W,Z)$.
Then $\gamma$ is Lorentzian. Furthermore, the null normal
$\ell$ to $\hat S_{p}$ projects to a null vector in the
quotient, as all Killing vectors are tangent to $\hat S_{p}$.
So $\hat S_p/ \T^{s-r}$ is a smooth null hypersurface through
the projection $q_0$ of $p_0$ by the quotient map. We continue
to denote by $K_\kl i$, $i=1,\ldots,r$, the Killing vectors of
$(\mcM/ \T^{s-r},\gamma)$ generating the remaining $\T^r$
action. Then the $K_\kl i$'s, $i=1,\ldots r$ are commuting
Killing vectors vanishing at $q_0$. In normal coordinates,
after perhaps redefining the $K_\kl i$'s if necessary, the
matrices $\nabla_\mu (K_\kl i)_\nu|_{q_0}$ can be represented
by consecutive two-by-two blocks on the diagonal, with the
associated non-trivial invariant spaces being spacelike. It
follows that
$$
 n+1-(s-r)=\dim \mcM/ \T^{s-r}\ge 2r+1
 \;,
$$
where the ``$+1$" at the right-hand-side accounts for at least
one timelike direction. Since $n+1=s+3$ by hypothesis, we
obtain
$$3+r\ge 2r+1
 \;,
$$
hence $r=1$ or $2$. Since we are assuming that we are at an
intersection point of axes, $r$ equals to two. Then
$s-r=n-2-2=n-4$, and dim $\mcM/ \T^{s-r}=n+1-(s-r)=5$. This
shows that the subspace of $T_{q_0}\mcM/ \T^{s-r}$ invariant
under $\T^r$ is one-dimensional timelike. But the normal vector
at $q_0$ to $\hat  S_{p}/ \T^{s-r}$ is a null vector invariant
under the action of $\T^{r}$. We conclude that $\hat S_{p}$
cannot pass through an intersection point of the axes. The
remaining arguments of the proof of \cite[Theorem~5.6]{ChCo}
apply now without modification.
\qed

\section{The area function near the axis}
 \label{sSergo}

 In this section we prove:

\begin{Theorem}
 \label{TdocA}  Under the hypotheses of Theorem~\ref{TCdsc},
suppose moreover that $G=\R\times \T^{s }$ with
$s+1$--dimensional principal orbits, $0\le s\le n-2$. If the
orthogonal integrability condition \eq{intcond2} holds, then
$\,\Span\{\Kz,\ldots,K_\kl s\}$ is timelike throughout the
domain of outer communications.
\end{Theorem}

Note that the dimension of $\,\Span\{\Kz,\ldots,K_\kl s\}$ is \emph{not}
assumed to be constant.

Theorem~\ref{TdocA} generalises to higher dimensions the
\emph{Ergoset theorem} of~\cite{ChCo}.

\proof Positivity of $W$ on $\doc\setminus \mcA$, has already
been established in Theorem~\ref{Tdoc2}. Consider thus a point
$p\in \mcA\cap \doc$. It follows from
\cite[Corollary~3.8]{ChCo} that $\Kz$ is transverse to
$\,\Span\{K_\kl 1,\ldots,K_\kl s\}|_p$, so Theorem~\ref{TdocAx}
below, and the calculations there, apply. If
$\,\Span\{\Kz,\ldots,K_\kl s\}|_p$ is null,
Theorem~\ref{TdocAx}  shows that
$$
 \{q\in \mcM \ | \
 W(q)=0\}\cap (\doc\setminus \mcA) \ne \emptyset
 \;,
$$
which  is not possible by Theorem~\ref{Tdoc2}. On the
other hand, \eq{Wdetid1xc} below shows that a spacelike
$\,\Span\{\Kz,\ldots,K_\kl s\}|_p$   would lead to a negative
function $W$ at nearby points lying on geodesics orthogonal to
$\,\Span\{\Kz,\ldots,K_\kl s\}|_p$ , which is again not possible
by Theorem~\ref{Tdoc2}. \qed

\medskip

It remains to prove:

\begin{Theorem}
 \label{TdocAx} Let $n\ge 3$, and let $(\mcM,\fourg)$ be an
$(n+1)$--dimensional Lorentzian manifold with an effective action of
$\R\times \T^{s}$
 \ptc{condition on s missing, I am not sure that this is deliberate}
by isometries satisfying the orthogonal integrability
condition \eq{intcond2}. Assume that the orbits of $\T^{s}$ are
spacelike,  that $Z(p)=0$ for some  $p\in \mcA$, and that $\Kz$
is  transverse to $\Span\{ K_\kl{ 1},\ldots, K_\kl s\}|_p$. If
 $\Span\{\Kz, \ldots, K_\kl s\}|_p$ is a null subspace of
 $T_p\mcM$, then  $W$ vanishes on
$$
 \mbox{\rm exp}_p(\Span\{\Kz, \ldots, K_\kl s\}|_p^\perp)
$$
which, for any
neighborhood $\mcU$ of $p$,  has a non-empty intersection with $\mcU\setminus \mcA$.
\end{Theorem}

\proof
%
Throughout this proof we shall interchangeably think of the $K_\kl \mu$'s
as elements of the Lie algebra of the group of isometries of
$(\mcM,\fourg)$, or as vector fields on $\mcM$.

Without loss of generality we can assume that the linearly independent
Killing vectors $K_\kl i$, $i=1,\ldots,s$, have $2\pi$--periodic orbits. By
hypothesis we have
\bel{acausalki}
\fourg(K_\kl i,K_\kl i) \ge 0\;, \quad \mbox{with} \quad  \fourg(K_\kl i,K_\kl i)|_q=0\quad \Longleftrightarrow \quad K_\kl i |_q =0
\;.
\ee
(Note that periodicity of orbits implies \eq{acausalki} in causal
space-times. In view of \eq{acausalki},  $Z(p)=0$ is only possible if some
linear combination of the $K_\kl i$'s vanishes at $p$, and then $W(p)=0$
as well. Thus
$$
 \mcA\subset \{p\in \mcM\ | \ W(p)=0\}
 \;.)
$$

Let $G_p\subset \T^s$  denote the connected component of the
identity of the set of $g\in \T^s$ which leave $p$ fixed; since
$Z(p)=0$ this is a closed non-trivial Lie subgroup of $\T^s$.
Hence $G_p=\T^r$ for some $0<r\le s$, and we can choose a new
basis of $\Span\{K_\kl 1,\ldots, K_\kl s\}$, still denoted by
$K_\kl i$, so that all $K_\kl i$'s remain $2\pi$--periodic, and
$K_\kl 1,\ldots,K_\kl r$ generate $\T^r$.

Since the isotropy group of $p$ has dimension $r$, the spacelike subspace
$$
 \Span\{K_\kl {r+1},\ldots,K_\kl s\}|_p
$$
of the tangent space at $p$ has dimension $s-r$, hence its orthogonal
$$
 \Span\{K_\kl {r+1},\ldots,K_\kl s\}|_p^\perp
$$
is a timelike subspace of
dimension $n+1+r-s$. The first space is  invariant under $\T^r$,  and so
must be the second.

Let $\hat T_p\in  \Span\{K_\kl {r+1},\ldots,K_\kl s\}|_p^\perp$ be any timelike vector at $p$, set
\bel{Tav}
 T_p:= \int_{\T^r} g_* \hat T_p \,dg \in \Span\{K_\kl {r+1},\ldots,K_\kl s\}|_p^\perp
 \;,
 \ee
where $dg$ is the translation-invariant measure on $\T^r$ normalised to
unit volume. Then $T_p$ is invariant under $\T^r$. Hence the  space
$T^\perp_p$   of vectors orthogonal to $T_p$ is also invariant under
$\T^r$. Multiplying $T_p$ by a suitable real, we can without loss of generality
assume that $T_p$ is unit, future directed.

 A standard argument (see, e.g.,~\cite[Appendix C]{ACD2}) shows
that for $i=1,\ldots, r$ each $K_\kl i$ vanishes on
$$
\mcA_{p,\kl i}:=\mexp_p(\Ker\, \nabla K_\kl i)
\;,
$$
and that
$\mcA_{p,\kl i}$ is totally geodesic. Note that $T_p\in \Ker\, \nabla K_\kl i$ for
$i=1,\ldots,r$, which implies that those $\mcA_{p,\kl i}$'s are timelike, and
that
$$
 \mcA_p:=\cap_{i=1}^r \mcA_{p,\kl i}
$$
is a non-empty totally geodesic timelike submanifold of $\mcM$ containing
$p$.

Since $[K_\kl \mu,K_\kl i]=0$ we have at $p$,  for $i=1,\ldots,r$ and for all $\mu$,
$$
K_\kl \mu^\alpha \nabla _\alpha K_\kl i =
K_\kl i^\alpha \nabla _\alpha K_\kl \mu = 0
 \;,
$$
so $K_\kl \mu \in \Ker \, \nabla K_\kl i$ for $i=1,\ldots,r$, and the arguments of
\cite[Proposition~C.1]{ACD2} show that each
$K_\kl \mu$ is tangent to all $\mcA_{p,\kl i}$'s as well.

Alternatively, since the Killing vectors commute, $\fourg(K_\kl i,K_\kl i) $ is
invariant under the flow of $K_\kl \mu$. So if $\fourg(K_\kl i,K_\kl i)$
vanishes at $p$, then it vanishes at $\phi_t[\Kz](p)$, where (as before) $\phi_t[K]$
denotes the flow of a Killing vector $K$; the vanishing of $K_\kl i$ at
$\phi_t[\Kz](p)$ follows then from \eq{acausalki}.

If $s=r$, we let $\hyp_\mcO=\mexp_p|_{\mcO}(T^\perp_p) $,
where $\mcO$ is any open neighborhood of $p$ lying within the
injectivity radius of $\mexp_p|_\mcO$ sufficiently small so that
$\hyp_\mcO$ is spacelike, while  $\mexp_p|_\mcO$ denotes the
exponential map centred at $p$ in the spacetime $(\mcO,g|_\mcO)$.

Otherwise we consider the intersection
$$
\sigma_p:=\Span\{K_\kl {r+1},\ldots,K_\kl s\}|_p^\perp\cap T_p^\perp\;,
$$
since $T_p \in \Span\{K_\kl {r+1},\ldots,K_\kl s\}|_p^\perp$, $\sigma_p$ is
 a spacelike, $(n+r-s)$--dimensional, subspace of the tangent space at
$p$, invariant under $\T^r$. Let $\mcO$ be a sufficiently small open
neighborhood of $p$ lying within the injectivity radius of $p$, then
\bel{Sigmadef}
 \Sigma:=\mexp_p(\sigma_p)\cap \mcO
\ee
is a smooth $(n+r-s)$--dimensional spacelike submanifold of $\mcO$ invariant
under $\T^r$.

Let $\hat G$ be the group generated by $\Span\{K_\kl {r+1},\ldots,K_\kl s\}$, and let $\hyp_\mcO$ denote the union of the orbits of $\hat G$, within
$\mcO$, passing through $\Sigma$. (Note that this reduces to the
previous definition when $r=s$.) Passing to a subset of $\mcO$ if
necessary, $\hyp_\mcO$ is then a smooth spacelike hypersurface in
$\mcO$ to which all Killing vector field $K_\kl i$, $i=1,\ldots,s$, are
tangent. Indeed, this is already so by construction for $i=r+1,\ldots,s$.
For the remaining $i$'s, let $T$ denote the field of future directed unit vectors normal to
$\hyp_\mcO$. Again by construction we have
$$
 \mcL_{K_\kl i} T= 0\;, \quad i=r+1,\ldots,s
 \;,
$$
where $\mcL$ denotes Lie derivation. This implies
$$
 \mcL _{K_\kl i}\Big(\fourg(K_\kl \mu, T)\Big)= 0\;, \quad i=r+1,\ldots,s
 \;,
$$
and since $\fourg(K_\kl j, T)= 0$   at $\Sigma$ for $j=1,\ldots,s$,  we
obtain, along $\hyp_\mcO$,
\bel{iperpeq}
 \fourg(K_\kl i,T)=0\;,\quad i=1,\ldots, s
 \;.
\ee

Now,  $\Kz$   is
transverse to $\hyp_\mcO$ by hypothesis (passing again to a subset of $\mcO$ if
necessary). Moving  $\hyp_\mcO$ with the flow of $\Kz$ we obtain a
function $t$, near $p$, defined by  setting
\bel{tfundef}
 t(p)=s \ \mbox{ iff } \ \phi_{-s}(p)\in \hyp_\mcO
 \;.
\ee
The function $t$ is a time-function, as notation suggests:
indeed, the level sets of $t$ are spacelike, which implies that
$\nabla t$ is timelike. Clearly $\hyp_\mcO=\{t=0\}$.Similarly
to the proof of \eq{iperpeq} along $\hyp_\mcO$, commutativity
of $\Kz$ with the $K_\kl i$'s shows that the $K_\kl i$'s are
tangent to the level sets of $t$.  Letting, away from $\hyp_\mcO$, $T$ be the field of
future directed unit vectors normal to the level sets of $t$,
\eq{iperpeq} holds now in a neighborhood of $p$.
\renewcommand{\changedX}{{\Kz}}%

We set
\bel{qwdef}
 w:=K_\kl 0^\flat \e\ldots \e K_\kl {s}^\flat
 \;, \qquad \hat w :=
 K_\kl 1^\flat \e\ldots \e K_\kl {s}^\flat
 \;,
\ee
where for any vector field $Y$ we set $Y^b:=\fourg(Y,\cdot)$. By definition
we have
 \bel{wdef2}
 w(\Kz,\ldots, K_\kl s) = -W
 \;,
 \qquad
 \hat w (K_\kl 1 ,\ldots, K_\kl s) = Z
 \;.
\ee

We need an equation of Carter~\cite{CarterJMP}:
\begin{equation}
\label{eqcarter}
  dW \e w = W dw\;.
\end{equation}
To prove \eq{eqcarter}, let $F=\{W=0\}$; note that the result is trivial on the
interior $\mathring F$ of $F$, if non-empty.  By continuity, it then suffices to
prove \eq{eqcarter} on $\mcM\setminus F$. So let $\mcU$ be the set of
points in $\mcM\setminus F$ at which the Killing vectors are linearly
independent. Consider any point $p\in \mcU$, and let $(x^a,x^A)$,
$a=0,\ldots,s$, be local coordinates near $p$ chosen so that $K_\kl a
=\partial_a $ and $\text{Span}\{\partial_a\} \perp\text{Span}\{ \partial_A\}$;
this is possible by \eq{intcond2}. Then
\bel{wdef} w = - W dx^0\wedge\ldots \wedge dx^{s}
 \;,
\ee
and \eq{eqcarter} follows near $p$, hence on
$\overline{\mcU}=\overline{\mcM\setminus F}$,  and hence everywhere.

Recall that $\changedX$ is causal at $p$, hence transverse to
$\hyp_\mcO$.  Passing to a subset of $\mcO$ if necessary, we redefine
$T$ to be the field of vectors normal to the level sets of the time function
$t$, as defined \eq{tfundef}, normalised so that $\fourg(T,\changedX)=1$;
the new $T$ is thus a smooth non-zero multiple of the previous one. Since
$\fourg(K_\kl i,T)=0$,
 \bel{wdef2xx}
 w(T,K_\kl 1,\ldots K_\kl s) =
 \Kz^\flat (T) (K_\kl 1^\flat \e\ldots \e K_\kl {s}^\flat)(K_\kl 1,\ldots, K_\kl s) = Z
 \;.
\ee

Let $\gamma$ be any affinely parameterised geodesic such that
$\gamma(0)=p$ and $\dot\gamma(0) \perp K_\kl \mu$ for all
$\mu=0,\ldots,s$; it is well known that then
\bel{kmuorth}
 \fourg(K_\kl \mu,\dot \gamma)=0
\ee
along $\gamma$. We then have by \eq{eqcarter}, \eq{wdef2xx} and
 \eq{kmuorth},
\bel{Wweq2} \underbrace{Z\frac{dW}{ds} }_{(dW\wedge w  )(\dot \gamma,
T, K_\kl 1,\ldots,K_\kl s)}=  W dw(\dot \gamma, T, K_\kl 1,\ldots,K_\kl s) \;.
\ee
Let $\alpha^\kl\mu$ denote the $s$--form obtained by omitting the $K_\kl
\mu$ factor in $ w$, and multiplied by $(-1)^\mu $. Similarly let $\beta^\kl i$
denote the $(s-1)$--form obtained by omitting the $K_\kl i$ factor in $
(-1)^i \hat w$. Using the summation convention on the index $\kl \mu$ we
have
\beal{intermx0} dw(\dot \gamma, T, K_\kl 1,\ldots,K_\kl s)
 & = &
(dK^\flat_\kl \mu \wedge  \alpha^\kl \mu)(\dot \gamma, T, K_\kl
1,\ldots,K_\kl s)
 \;,
\eea
Now,
\beal{intermx1}
 \phantom{xxx}
 (dK^\flat_\kl 0 \wedge  \alpha^\kl 0)(\dot \gamma, T, K_\kl
1,\ldots,K_\kl s)
 & = &
 dK^\flat_\kl 0 (\dot \gamma, T) \alpha^\kl 0(K_\kl
1,\ldots,K_\kl s)
 \\
 \nonumber
 & = &
 Z
 dK^\flat_\kl 0 (\dot \gamma, T)
 \;,
\eea
while, again summing over $\kl i$,
\beal{intermx}
 &&
 \\
 \nonumber
 (dK^\flat_\kl i \wedge  \alpha^\kl i)(\dot \gamma, T, K_\kl
1,\ldots,K_\kl s)
 & = &
 dK^\flat_\kl i (\dot \gamma, T) \alpha^\kl i(K_\kl
1,\ldots,K_\kl s)
 \\
 \nonumber
 &  &
 + \sum_j(-1)^j
 dK^\flat_\kl i (\dot \gamma, K_\kl j )\alpha^\kl i(\underbrace{T, K_\kl
1,\ldots,K_\kl s}_{\mathrm{no} \ K _\kl j})
\\
 \nonumber
 & = &
 dK^\flat_\kl i (\dot \gamma, T) \alpha^\kl i(K_\kl
1,\ldots,K_\kl s)
 \\
 \nonumber
 &  &
 + \sum_j(-1)^j
 dK^\flat_\kl i (\dot \gamma, K_\kl j )\beta^\kl i(\underbrace{K_\kl
1,\ldots,K_\kl s}_{\mathrm{no} \ K _\kl j})
 \;.
\eea
Using
 $i_{K_\kl j}d\hat w = \mcL_{K_\kl j}\hat w -d(i_{K_\kl j}
\hat w)=-d(i_{K_\kl j} \hat w)$, as well as further similar equations that
follow from $\mcL_{K_\kl i}K_\kl j =0$, one has
\beaa
 d\hat w(\dot \gamma, K_\kl 1,\ldots,K_\kl s) &=& (-1)^s d\hat w(K_\kl 1,\ldots,K_\kl {s },\dot \gamma)
 \\&=&
 (-1)^s  i_{\dot \gamma} i_{K_\kl s}\ldots i_{K_\kl 1} d\hat w
 \\&=&
- (-1)^s i_{\dot \gamma} i_{K_\kl s}\ldots i_{K_\kl 2} d( i_{K_\kl 1} \hat w)
=\ldots
 \\&=&
 -  i_{\dot \gamma}d(  i_{K_\kl s}\ldots i_{K_\kl 2} i_{K_\kl 1} \hat w)
 \\&=&
 -\frac{dZ}{ds} \;.
\eeaa
On the other hand,
\beaa
 d\hat w(\dot \gamma, K_\kl 1,\ldots,K_\kl s)
 &=&
 (dK^\flat_\kl i \wedge  \beta^\kl i)(\dot \gamma, K_\kl
1,\ldots,K_\kl s)
 \\
 & = &
  \sum_j(-1)^j
 dK^\flat_\kl i (\dot \gamma, K_\kl j )\beta^\kl i(\underbrace{K_\kl
1,\ldots,K_\kl s}_{\mathrm{no} \ K _\kl j}) \;.
\eeaa
Comparing with \eq{intermx}, we conclude
\beal{intermx3}
 \phantom{xx}(dK^\flat_\kl i \wedge  \alpha^\kl i)(\dot \gamma, T, K_\kl
1,\ldots,K_\kl s)
 & = &
 dK^\flat_\kl i (\dot \gamma, T) \alpha^\kl i(K_\kl
1,\ldots,K_\kl s)
 - \frac {dZ}{ds}
 \;.
 \nonumber
\eea
Collecting all this, we obtain our key equation:
\bel{Wweq3} \frac{d}{ds}\left(\frac W{Z} \right)= \underbrace{
 \Big(Z^{-1}\alpha^\kl i(K_\kl
1,\ldots,K_\kl s)
 dK^\flat_\kl i  +d\Kz^\flat\Big) (\dot \gamma, T) }_{=:f} \times \frac
W{Z} \;. \ee

We shall  need the following:

\begin{Lemma}
 \label{Lstrucor} Let $(\mcM,\fourg)$ be an $(n+1)$--dimensional
Lorentzian manifold with an effective action of $\R\times \T^{s}$ by
isometries.
 \ptc{condition on s}
Suppose that $\Kz$ is causal at $p$   while
 $\Span\{K_\kl i\}_{i=1}^s|_p$ is spacelike, and that the isotropy group of $p$ is
$\T^r$.
Then
\bel{rres}
 0\le r \le n-s
 \;,
\ee
and if $r>0$ there exist coordinates $(x^i,y^i,z^a)$,
$i=1,\ldots,r$,   and a basis $\{K_\kl i\}_{i=1}^s$, consisting
of $2\pi$ periodic Killing vectors, of the Lie algebra of $\T^s$
such that
\bel{canfo}
 K_\kl i = 
 ( x^i\partial _{y^i} - y^i\partial_{x^i})
 \;,
 \
 i=1,\ldots, r
 \quad \mbox{(no summation over $i$)}
  \;.
\ee
Furthermore, setting
\bel{rhorhodef}
 \rho_\kl i:= \sqrt{(x^i)^2+(y^i)^2}\;, \qquad \rho:= \sqrt{\rho_\kl
1^2+\ldots+ \rho_\kl r^2}
 \;,
 \ee
there exists a
constant $C$ such that we have, for all sufficiently small $\rho_\kl i$ and
$\rho_\kl j$,
\beal{metco} &
 \forall \  i=1,\ldots,r \quad
 C^{-1} \rho_\kl i^2\le \fourg(K_\kl i,K_\kl i)\le C \rho_\kl i^2
  \;,
  &
  \\& \forall \  i=1,\ldots,r\;,\ \forall \ j = {r+1,\ldots,s} \quad
 \fourg(K_\kl i,K_\kl j) \le C \rho_\kl i \rho_\kl  j \rho
  \;,
  &
  \\
  &
   \phantom{xxx}
 \forall \  i=1,\ldots,r\;,\ \forall \ \mu \in\{0,r+1,\ldots,s\} \; \quad
  \fourg(K_\kl i,K_\kl\mu) \le C\rho_\kl i   \rho
  \;.
  &
\eeal{metco2}
\end{Lemma}

\proof
Let $\{\tilde K_\kl i\}_{i=1,\ldots,s}$ denote any basis of the
Lie algebra of $\T^s$, formed by $2\pi$--periodic Killing
vector fields.  Let $\{\hat K_\kl i\}_{i=1,\ldots,r}$ be any
basis of the Lie algebra of $\T^r$, again formed by
$2\pi$--periodic Killing vector fields. We can complete $\hat
K_\kl i$ to a basis $\{\hat K_\kl i\}_{i=1}^s$ of the Lie
algebra of $\T^s$ using the $\tilde K_\kl i$'s, and we set
$\hat K_\kl 0=\Kz$.

By construction, the manifold $\Sigma$ defined by \eq{Sigmadef}, is  a
smooth $(n-s+r)$--dimensional spacelike submanifold  of $\mcM$
transverse at $p$ to  the $\hat K_\kl i$'s, $i\in\{ r+1,\ldots ,s\}$ \emph{and}
to the vector  $T_p$ of \eq{Tav}. Let
$$
 \mcU\subset \Sigma
$$
be a sufficiently small coordinate ball around $p$. Let, as before, $\hat G$
be obtained by exponentiating $\Span\{K_\kl {r+1},\ldots,K_\kl s\}$, and let
$\mcV$ be the union of the orbits of $\hat G$ passing through $\mcU$.
Passing to a subset of $\mcU$ if necessary, we can without loss of
generality assume that  the action of $\R\times \T^{s-r}$ generated by the
$\hat K_\kl \mu$'s, with $\mu\in\{0,r+1,\ldots, s\}$, on $\mcV$ is free, and
by elementary considerations one obtains
$$
 \mcV=\mcU\times \R \times \T^{s-r}
 \;.
$$

We note that the function $t$ of \eq{tfundef} defines
a unique $\T^s$--invariant time function on $\mcV$, so that we have proved:

\begin{Proposition}
 \label{Psnp}
 Under the hypotheses of Theorem~\ref{TdocA}, there exists an $\R\times
 \T^s$--invariant stably causal neighborhood of $p$.
\qed
\end{Proposition}

(We note that some considerations so far could have been
considerably simplified if the conclusions of
Proposition~\ref{Psnp} have been known a priori, by averaging
any time function as in the proposition over $\T^s$.)

Returning to the proof of Lemma~\ref{Lstrucor}, let $\hthreeg$ be the
metric induced on $\hyp_\mcO$ by $\fourg$. Then $\hthreeg$ is a
Riemannian metric invariant under $\T^r$. Let $\gamma$ denote the
orbit-space metric on $\mcU$,
\bel{gammetx}
 \forall \ X,Y\in T\mcU\qquad
 \gamma(X,Y)= \hthreeg(X,Y) - h^{\kl i \kl j}\hthreeg(X,K_\kl i) \hthreeg(Y,K_\kl j)
 \;,
 \ee
where $h^{\kl i\kl j}$ denotes the matrix inverse to $\hthreeg(K_\kl i,K_\kl
j)$, $i,j=r+1,\ldots,s$, and  in \eq{gammetx} one sums over $i,j$ in the last
range. It is simple to check,
using the Cauchy-Schwarz inequality, that
$\gamma$ is Riemannian, so that the group $\T^r$ acts locally on the
Riemannian manifold $(\mcU,\gamma)$ by isometries, with complete
orbits near $p$. We infer that near $p$ the original orbit space $\mcM/(\R\times \T^s)$ is
diffeomorphic to $\mcU/\T^r$.

 Now, $\T^r$ acts effectively on $(T_p\mcU,\gamma|_p)$ by
isometries, so we can view $\T^r$ as a closed abelian subgroup of
$SO(n-s+r)$, such that the principal orbits of the action of $\T^r$ on
$\R^{n-s+r}$ are $r$-dimensional.

Let $G\subset SO(n-s+r)$ denote any maximal torus containing $\T^r$,
by~\cite[Theorem~16.2]{Bump} $G$ is conjugated to a standard maximal
torus as in~\cite[Example~6.21]{Fegan}, hence
\bel{rcalc}
r \le \dim G = \lfloor \frac {n-s+r} 2 \rfloor \le   \frac {n-s+r} 2   \quad \Longrightarrow \quad 0\le r\le n-s
\;.
\ee

Consider the simplest case $r=1$,  then $\dim \Ker \nabla K_\kl 1 =
n+1-2=n-1$.
Let $ (x^1,y^1,z^a)\equiv (x^A,z^a)$   be the coordinates
of~\cite[Proposition~C.1]{ACD2} (denoted by $(x^A,x^a)$ there, and
constructed there under the assumption that the metric is Riemannian, but
the result holds for a Lorentzian $\fourg$ whenever $\Ker \nabla X$
contains a timelike vector), with $n$ there replaced by $n+1$, $X$ there
equal to $K_\kl 1$, and $\ell $ there equal to one. The lemma follows now
from~\cite[Equation~(C.8)]{ACD2}:
\beal{metstrucor}
 \fourg  &= & \sum_{i=1}^\ell\left( (dx^i)^2 + (dy^i)^2\right)+\sum_{A,B} O(\rho^2)dx^A dx^B+\sum_{A,a} O(\rho)dx^A dz^a
 \\
 &&  + \fourg_{ab}|_{\rho=0}dz^a d z^b +\sum_{a,b} O(\rho^2)  dz^a dz^b
 \;,
\eean
where $\rho^2=\rho_\kl 1^2 + \ldots +\rho_\kl \ell ^2 $.

In general, by the already mentioned~\cite[Theorem~16.2]{Bump}
and~\cite[Example~6.21]{Fegan}, there exists an orthonormal basis of
$T_p M$ so that the flows of $\hat K_\kl i$  on $T_p M$, $i=1,\ldots, r$,
are generated by  linear combinations of vector fields $K_\kl 1$ and
$K_\kl 2$ as in \eq{canfo}. Equivalently,  the $K_\kl i$'s,   $i=1,\ldots, r$,  take the
form \eq{canfo} near $p$ in the associated normal coordinates centred at $p$.
Applying~\cite[Proposition~C.1]{ACD2} to
$$
 X=K_\kl 1+ \ldots + K_\kl r
  \;,
$$
with $\ell $ there equal to $r$, our claims follow again
from~\eq{metstrucor}.
%
\qed

\bigskip

For further reference we note the following variation of
Lemma~\ref{Lstrucor}, with essentially identical, but somewhat simpler,
proof:

\begin{Lemma}
 \label{Lstrucorx} Let $(M,\hthreeg)$ be an $n$--dimensional
Riemannian manifold with an effective action of $\T^s$ by
isometries.. If $\T^r$ is the isotropy group of $p$, then
\eq{rres} holds, and for $r>0$ there exist coordinates
$(x^i,y^i,z^a)$, $i=1,\ldots,r$,   and a basis $\{K_\kl
i\}_{i=1}^s$, consisting of $2\pi$ periodic Killing vectors, of
the Lie algebra of $\T^s$ such that \eq{canfo} holds.
Furthermore, letting $\rho$ and $\rho_\kl i$ be as in
\eq{rhorhodef}, there exists a constant $C$ such that we have,
for all sufficiently small $\rho_\kl i$ and $\rho_\kl j$,
\beal{metcox} &
 \forall \  i=1,\ldots,r \quad
 C^{-1} \rho_\kl i^2\le \hthreeg(K_\kl i,K_\kl i)\le C \rho_\kl i^2
  \;,
  &
  \\& \forall \  i\ne j\in\{1,\ldots,r\} \quad
\hthreeg(K_\kl i,K_\kl j) \le C \rho_\kl i \rho_\kl  j \rho
  \;,
  &
  \\
  &
   \phantom{xxx}
 \forall \  i=1,\ldots,r\;,\ \forall \ j = {r+1,\ldots,s}, \; \quad
  \hthreeg(K_\kl i,K_\kl j) \le C\rho_\kl i   \rho
  \;.
  &
\eeal{metco2x}
\qed
\end{Lemma}

\medskip

We return now to the analysis of \eq{Wweq3}. The case $s=1$ has already
been covered in~\cite{ChCo}, so we assume $s\ge 2$. Set
$$
 \check Z_\kl r:= \det \left(\fourg(K_\kl \mu, K_\kl \nu)_{\mu,\nu=0,r+1,\ldots,s}\right)
 \;,\quad
 Z_\kl r:= \det \left(\fourg(K_\kl i, K_\kl j)_{i,j=r+1,\ldots,s}\right)
 \;.
 $$
\renewcommand{\hthreeg}{\fourg}%
Suppose, first, that $k=1$. Then, after exchanging the zeroth and first  row,
and then the zeroth and first  column, $W$ is minus the determinant of a
matrix of the form
\bel{Wmat}
 \left(
 \begin{array}{cccc}
    \hthreeg(K_\kl 1,K_\kl 1)    & O(\rho^2) & \ldots & O(\rho^2) \\
   O(\rho^2) & \star & \ldots & \star \\
   \vdots &  \vdots & \ddots & \vdots \\
    O(\rho^2) & \star & \ldots & \star
 \end{array}
\right)
\;.
\ee
\Eq{Wmat}, and a similar equation for $Z$, leads to
\bel{Wdetid}
 W= -\hthreeg(K_\kl 1,K_\kl 1)  \check   Z_\kl 1 + O(\rho^{4})\;,
 \quad
 Z=  \hthreeg(K_\kl 1,K_\kl 1)  Z_\kl 1(1 + O(\rho^{2}))
 \;,
\ee
(recall that $Z_\kl 1(p)$ does not vanish by hypothesis).  Using \eq{metco} we
conclude that  $W/Z$ approaches $-\check Z_\kl 1/ Z_\kl 1 $ as one approaches
$p$ along $\gamma$.

Next, we wish to show that the function $f$ defined in \eq{Wweq3} is bounded; this requires an analysis of the term
\beal{Zft}
&&
 \\
 \nonumber
 {Z^{-1}\alpha^\kl i(K_\kl
1,\ldots,K_\kl s)\;
 dK^\flat_\kl i  (\dot \gamma, T)}
 &=&
Z^{-1}\alpha^\kl 1(K_\kl
1,\ldots,K_\kl s)\;
 dK^\flat_\kl 1 (\dot \gamma, T) +
 \\
 && \sum_{i>1}
Z^{-1}\alpha^\kl i(K_\kl
1,\ldots,K_\kl s)\;
 dK^\flat_\kl i  (\dot \gamma, T)
\;.
\eean
Writing $h_{\mu\nu}$ for $\hthreeg(K_\kl \mu,K_\kl \nu)$,  by definition we have
\beaa
&
 \alpha^\kl 1(K_\kl
1,\ldots,K_\kl s) = \epsilon^{i_0 i_2\ldots i_s} h_{0i_0}h_{2i_1}\ldots h_{si_s} = O(\rho^2)
 \;,
 &
 \\
 &
i>1: \quad  \alpha^\kl i(K_\kl
1,\ldots,K_\kl s) = (-1)^i \epsilon^{i_0 i_1\ldots i_s} h_{0i_0}
 \underbrace{h_{1i_1}\ldots h_{si_s} }_{\mbox{\scriptsize no} \ h_{i j_i } \ \mbox{\scriptsize factor}}= O(\rho^2)
 \;,
 &
\eeaa
and boundedness of $f$ readily follows.

When $r=2$ we set $\rho^2=\sqrt{\rho_\kl 1^2 + \rho^2 _\kl 2}$;
  then, after moving the zeroth row past the next two ones,
  similarly for the zeroth column, $W$ is the minus the
  determinant of
\bel{Wmat2}
 \left(
 \begin{array}{ccccc}
 \hthreeg(K_\kl 1,K_\kl 1) & O(\rho_\kl 1 \rho_\kl 2 \rho)& O(\rho_\kl 1  \rho)   & \ldots & O(\rho_\kl 1  \rho)   \\
  O(\rho_\kl 1 \rho_\kl 2 \rho) & \hthreeg(K_\kl 2,K_\kl 2) & O(  \rho_\kl 2 \rho)  & \ldots & O(  \rho_\kl 2 \rho)  \\
  O(\rho_\kl 1  \rho) &  O(  \rho_\kl 2 \rho)  & \star & \ldots & \star \\
  \vdots & \vdots & \vdots & \ddots & \vdots \\
 O(\rho_\kl 1  \rho) & O(  \rho_\kl 2 \rho)  & \star & \ldots & \star
 \end{array}
\right)
\;.
\ee
One finds
\beal{Wdetid1}
 W&=&  -\hthreeg(K_\kl 1,K_\kl 1) \hthreeg(K_\kl 2,K_\kl 2) \check Z_\kl 2 + O(\rho_\kl 1^2 \rho_\kl 2^2 \rho^2)\;,
 \\
 Z&=& \hthreeg(K_\kl 1,K_\kl 1) \hthreeg(K_\kl 2,K_\kl 2)  Z_\kl 2(1 + O (  \rho^2))\;,
\eeal{Wdetid2}
The form of the error terms  plays a key role when
taking the quotient $W/Z$ below, so it deserves a more careful
justification. We start with the determinant
$Z$:
$$
 Z=\epsilon^{i_1 \ldots i_s} h_{1i_1}\ldots h_{si_s}
 \;.
$$
Let us write
$$
 \epsilon^{i_1 \ldots i_s} _{i_1\ne 1}h_{1i_1}\ldots h_{si_s}
$$
for a sum where $i_1$ is not allowed to take the value one, and
$$
 \epsilon^{i_1 \ldots i_s}_{i_1\ne 1, i_2\ne 2} h_{1i_1}\ldots h_{si_s}
$$
for a sum where $i_1$ is not allowed to take the value one and $i_2$ is
not allowed to take the value two, etc. Then
\beaa
 Z& =&h_{1 1}\epsilon^{1 i_2 \ldots i_s}  h_{2i_2}\ldots h_{si_s}+ \epsilon^{i_1 \ldots i_s} _{i_1\ne 1} h_{1i_1}\ldots h_{si_s}
 \\
 &= &  \underbrace{h_{1 1}h_{22}\epsilon^{1 2 i_3 \ldots i_s} h_{3i_3} \ldots h_{si_s}}_{I}
  + \underbrace{h_{1 1}\epsilon^{1 i_2 \ldots i_s}  _{i_2\ne 2}h_{2i_2} \ldots h_{si_s}}_{II}
  \\
   &&
  + \underbrace{h_{22} \epsilon^{i_1 2 i_3\ldots i_s}  _{i_1\ne 1}h_{1i_1} h_{3i_3}\ldots h_{si_s}}_{III}
  +  \underbrace{\epsilon^{i_1 \ldots i_s} _{i_1\ne 1, i_2\ne 2} h_{1i_1}\ldots h_{si_s}}_{IV}
 \;.
\eeaa
The term $I$ is the main term $ \hthreeg(K_\kl 1,K_\kl 1) \hthreeg(K_\kl
2,K_\kl 2)  Z_\kl 2$  in \eq{Wdetid2}. In each term of the sum $II$ one of the
indices $i_k\ne i_2$ has to be a two, so each term in that sum contains a
factor $h_{2i_2}h_{i_k2}=O(\rho_\kl 2^3 \rho )=O(\rho_\kl 2^2 \rho^2)$. Taking into account
$|h_{11}|\le C\rho_\kl 1^2$ we obtain $|II|\le C \rho_\kl 1^2 \rho_\kl 2^2
\rho^2$, which can be factored as $h_{11}h_{22} Z_\kl 2 O(\rho^2)$. The
estimate on $III$ follows by symmetry, the analysis of $IV$ proceeds along
the same lines.

It should be clear from \eq{Wmat2} that the calculation for $W$ is identical,
after grouping $\Kz$ with the $K_\kl i$'s, $i=3,\ldots, s$. The only difference
is in the last step, where we cannot factor out $\check Z_2$, as we are
allowing it to vanish at $p$.

Without much further effort, the reader should be able to   conclude that for all $r$
\beal{Wdetid1xc}
 W&=&  -\hthreeg(K_\kl 1,K_\kl 1) \cdots \hthreeg(K_\kl r,K_\kl r) \check Z_\kl r + O(\rho_\kl 1^2 \ldots \rho_\kl r^2 \rho^2)\;,
 \\
 Z&=& \hthreeg(K_\kl 1,K_\kl 1) \cdots \hthreeg(K_\kl r,K_\kl r) Z_\kl r (1 + O (  \rho^2))\;,
\eeal{Wdetid2xc}
so that
\bel{limit}
 \lim_{s\searrow0} \frac WZ (\gamma(s)) =-\frac {\check Z_\kl r}{Z_\kl r} \Big|_p
 \;, \quad \mbox{ with }\ Z_\kl r(p)\ne 0
 \;.
\ee
It follows that the quotient $W/Z$ has a vanishing limit at $p\in\mcA$ either if $\Kz \in
\Span\{K_\kl {r+1},\ldots,K_\kl s\}|_p$, or if
 $\Span\{\Kz,K_{\kl {r+1}},\ldots,K_\kl s\}|_p$ is a null subspace of
$T_p\mcM$. The former possibility does not occur since $\Kz$ is
transverse  to $\Span\{K_\kl {r+1},\ldots,K_\kl s\}|_p$ by hypothesis.

Again  for $r=2$, consider the function $f$ of \eq{Wweq3}, the not-obviously-bounded
part of which we write now as
\beal{Zftx}
&&
 \\
 \nonumber
 {Z^{-1}\alpha^\kl i(K_\kl
1,\ldots,K_\kl s)\;
 dK^\flat_\kl i  (\dot \gamma, T)}
 &=&
Z^{-1}\alpha^\kl 1(K_\kl
1,\ldots,K_\kl s)\;
 dK^\flat_\kl 1 (\dot \gamma, T) +
 \\
 &&
Z^{-1}\alpha^\kl 2(K_\kl
1,\ldots,K_\kl s)\;
 dK^\flat_\kl 2  (\dot \gamma, T)+
 \nonumber
 \\
 && \sum_{i>1}
Z^{-1}\alpha^\kl i(K_\kl
1,\ldots,K_\kl s)\;
 dK^\flat_\kl i  (\dot \gamma, T)
\;.
\eean
By definition we have, for $s\ge 3$ (the calculation for $s=2$ is
typographically different, but otherwise identical),
\beaa
 \alpha^\kl 1(K_\kl
1,\ldots,K_\kl s) & =  & \epsilon^{i_0i_2\ldots i_s} h_{0i_0}h_{2i_1}\ldots h_{si_s}
 \\
  & =  & \underbrace{h_{22}}_{O(\rho_\kl 2^2)} \epsilon^{i_0 2 i_3 \ldots i_s} h_{0i_0} h_{3i_3}
   \ldots h_{si_s}
  +    \epsilon^{i_0i_2\ldots i_s}_{i_2 \ne 2} h_{0i_0}\underbrace{h_{2i_2}}_{O(\rho_\kl 2 \rho)}\ldots h_{si_s}
  \\
  & = &
  O(\rho_\kl 1 \rho_\kl 2^2 \rho )
 \;,
\eeaa
because each term in each of the sums above contains a factor $h_{r1}$,
$r\ne 1$, which is  $O(  \rho_\kl 1\rho )$; furthermore, one of the indices in
the second sum has to be equal to two, which gives a further factor
$O(\rho_\kl 2 \rho)$ in the second sum. One similarly obtains
\beaa
&
 \alpha^\kl 2(K_\kl
1,\ldots,K_\kl s) = \epsilon^{i_0 i_1 i_3\ldots i_s} h_{0i_0}h_{1i_1} h_{3i_3}\ldots h_{si_s} = O(\rho_\kl 1^2 \rho_\kl 2 \rho )
 \;,
 &
 \\
 &
j>2: \quad  \alpha^\kl j(K_\kl
1,\ldots,K_\kl s) = (-1)^j  \underbrace{\epsilon^{i_0 \ldots i_s} }_{\mbox{\scriptsize  no}\  i_j \ \mbox{\scriptsize index}} h_{0i_1}
 \ldots h_{si_s}= O(\rho_\kl 1^2 \rho_\kl 2^2 )
 \;
 &
\eeaa
Now, this does not suffice for estimating a quotient by $Z\approx \rho_\kl 1
^2 \rho_\kl 2^2$ for the first two terms in \eq{Zftx}. However, the missing
powers of $\rho_\kl 1$ and of $\rho_\kl 2$ are provided by $dK^\flat_\kl 1
(\dot \gamma, T) $ and $dK^\flat_\kl 1 (\dot \gamma, T) $:
$$
dK^\flat_\kl i (\dot \gamma, T) =  O(\rho_\kl i )
 \;.
$$
Indeed, $dK^\flat_\kl i (\dot \gamma, T)=0$ at $\{\rho_\kl i = 0\}$ for $i=1,\ldots, r$, and a
Taylor expansion of order zero near $\{\rho_\kl i=0\}$ gives the estimate.

Summarising, both for $r=1$ and $r=2$,
we have shown that the function $f$ defined in
\eq{Wweq3} is bounded along $\gamma$ near $p$.
A very similar analysis applies for higher $r$.

Now, if $\check Z_\kl k=0$ at $p$, then the limit at $p$ of $W/Z$ along
$\gamma$ vanishes by \eq{limit}. Using uniqueness of solutions of ODE's,
it follows from \eq{Wweq3} that $W$ vanishes along $\gamma$. To  finish
the proof it suffices to notice that any $\gamma$ with, e.g.,
$\dot\gamma(0)$ lying in the $(x^1,y^1)$ plane of the coordinates of
Lemma~\ref{Lstrucor} immediately leaves $\mcA$.
\qed

\section{Uniqueness of static solutions and zeros of Killing vectors}
 \label{Szokv}
\renewcommand{\changedX}{X}%

As pointed out in~\cite{ChCo}, the proof of uniqueness  of
higher dimensional globally hyperbolic, static, vacuum black
holes containing an asymptotically flat hypersurface, of
positive energy type, with boundary contained away from the
domain of outer communications, requires excluding zeros of the
Killing vector on degenerate components of the event horizon.
Our aim in this section is to prove that such zeros cannot
occur, as needed for the argument in~\cite{ChCo}.

We start with the following result, pointed out to us by
Abdelghani Zeghib, which is apparently well known among
researchers acquainted with hyperbolic geometry. For
completeness we provide the proof, as explained to us by
Zeghib:

\begin{Proposition}
 \label{PGhani}
Let $\changedX$ be a non-trivial Killing vector, and suppose
that $\changedX$ vanishes at $p\in \mcM$. Then there exists a
normal coordinate system $(x^\mu)$ near $p$ such that:
 \begin{enumerate}
\item either there exist constants $\beta_\mu\in \R$,
    $\mu=0,\ldots, m \le n/2$, not all zero, such that
\bel{Xfcase} \changedX = \beta_0
(x^0\partial_1+x^1\partial_0) + \sum_{i=1}^m \beta_i
(x^{2i+1}\partial_{2i}-x^{2i}\partial_{2i+1})    \;,
\ee
\item or there exists constants $a\in \R^*$ and
    $\beta_i\in\R$, $i=0,\ldots, m \le (n-1)/2$ such that
\bel{Xfcase2} \changedX = a \left(
(x^0-x^2)\partial_1+x^1(\partial_0+\partial_2) \right) +
\sum_{i=1}^m \beta_i
(x^{2i+1}\partial_{2i+2}-x^{2i+2}\partial_{2i+1}) \;.
\ee
\end{enumerate}
\end{Proposition}

\begin{Remarks}{\rm
1.
 Recall that every orthochronous, orientation preserving
 Lorentz matrix is the exponential of a matrix
 $\lambda^\mu{}_\nu= \partial_\nu X^\mu$, where $X^\mu$ is a
 Minkowski space-time Killing vector vanishing at the origin. So
 \eq{Xfcase}-\eq{Xfcase2} can also be used to obtain a canonical
 representation for Lorentz matrices.

 2. The coordinates of \eq{Xfcase} are unique, but those of \eq{Xfcase2} are not.
 }
\end{Remarks}
\begin{proof}
 Let $\lambda = \nabla
\changedX|_p$; in other words, $\lambda^\mu{}_\nu:=\nabla_\nu
     \changedX^\mu|_p$. Let $e_a$ be any ON frame at $p$ with
     $e_0$--timelike. Let $(x^\mu)$ denote the associated
     normal coordinates centered at $p$.  It is well known, and
     in any case note very difficult to show using the fact
     that isometries map geodesics to geodesics, that
     $\changedX=\lambda^\mu{}_\nu x^\nu \partial_\nu$. So to
     prove the result we need to classify the possible matrices
     $\lambda$, up to choice of ON-basis.

      Suppose that $\sigma\in \C\setminus \R$ is a root of the
     characteristic polynomial of $\lambda$, let $u+\ii v\in
     T_p\mcM\oplus \ii T_p \mcM$ be the corresponding
     eigenvector. Keeping in mind that one-dimensional
     eigenspaces lead to real eigenvalues, the space
     $\Span\{u,v\}$ is a two-dimensional space invariant under
     $\lambda$.  We claim that $\Span\{u,v\}$ is not null:
     otherwise it would contain a unique null direction, which
     would have to be mapped into itself by all the isometries
     $\exp (t\lambda)$. This would imply that $\Span\{u,v\}$   contains an
     eigenvector of $\lambda$ with real eigenvalue,
     contradicting $\sigma \in \C\setminus \R$. Thus,
     $\Span\{u,v\}$ is either a) timelike or b) spacelike.

     In the latter case b) we choose $e_0$ and $e_1$ so that
     $\Span\{u,v\}=\Span\{e_0,e_1\}$. Then the space
     $\Span\{u,v\}^\perp$ is a complementing spacelike subspace
     of $T_p\mcM$, invariant under $\lambda$, and we have
     reduced the problem to a Riemannian one, in dimension
     smaller by two.

     In the former case a) we pass to an ON basis of $T_p\mcM$ so
     that $\Span\{u,v\}=\Span\{e_{n-1},e_n\}$. Then  the space
     $\Span\{u,v\}^\perp$ is a complementing timelike subspace
     of $T_p\mcM$ invariant under $\lambda$, and we have
     reduced the dimension by two.

Note that if at any stage of this dimension-reduction
     process the metric becomes Riemannian, then the iteration
     of the argument in the last paragraph provides a finite number of
     two-dimensional orthogonal invariant spaces plus a
     Riemannian space, say $E$, invariant under $\lambda$, with all
     $E$--eigenvalues of $\lambda$  real.

     Now, generally, since
     $\lambda_{\mu\nu}$ is anti-symmetric we have
     $$
     0=\lambda_{\mu\nu}u^\mu u ^\nu = \sigma u_\mu u^\mu
     \;,
     $$
     which shows that $u$ is null unless $\sigma =0$. So on
     each timelike or spacelike one-dimensional eigenspace the
     action of the flow of $\changedX$ is trivial. Hence, if a
     Riemannian metric is obtained after any of the
     dimension-reduction steps described in this proof, after a
     finite number of further steps we obtain a basis where
     $\changedX$ takes the form \eq{Xfcase}.

     Iterating, we can decompose $T_p\mcM$ as an
     orthogonal sum of invariant two-dimensional spaces plus an
     invariant
     remainder, say again $E$. If $\lambda$ vanishes on $E$, then $\changedX$ takes the form
     \eq{Xfcase}, and we are done.

     Otherwise $\lambda$ maps a Lorentzian $E$ to $E$, we shall still
denote by $\lambda$
     the resulting map. By construction all roots of the
     characteristic polynomial of $\lambda|_E$ are real. Let
     $\sigma$ be such a root, and let $u$ be the corresponding
     vector. If $u$ is timelike or spacelike, then
     $\Span\{u\}^\perp$ is a complementing invariant space, and
     we can further reduce the dimension by splitting off
     $\Span\{u\}$ from $E$, and renaming the new space $E$. We
     continue in this way until there are, in $E$, no
     eigenvectors which are timelike or null. In particular $E$
     has no proper Riemannian eigenspaces.

Again, if at some stage one of $u$'s is timelike, we are in the
case \eq{Xfcase}.

      So, there eventually remains a space $E$ invariant under $\lambda$,
     with $\lambda$ having  only real eigenvalues, and only
     null eigenvectors. Suppose that there exist two such
     eigenvectors, $u$ and $v$, then $\Span\{u,v\}$ is
     timelike, invariant under the flow of $X$, with the
     complementing space Riemannian, or trivial. We avoid a contradiction with the fact
     that $\lambda$, restricted to $E$, has no proper
     Riemannian eigenspaces only if $\dim E=2$, leading to \eq{Xfcase}, and the proof is complete in this case.

     Otherwise $E$ contains only one null eigenvector $u$, and no
     invariant subspaces which are timelike or spacelike. The
     space $ \Span\{ u\}^\perp$ is a null subspace of $E$
     invariant under $\lambda$. Let $\{u,e_i\}$ be a basis of
     $\Span\{u\}^\perp$, then the $e_i$'s are necessarily
     spacelike. There exists a matrix $\alpha_i{}^j$ and
     numbers $\alpha_i$ such that
     $$
      \lambda e_i = \alpha_i u +\alpha_i{}^j e_j
\;.
$$
The numbers $(\alpha_i)$ behave as a vector under rotations of
$\Span\{e_1,\ldots\}$, so we can choose a rotation matrix
$\omega_i{}^j$ so that in the new basis $\hat e_i=\omega_i{}^j
e_j$ we have
     $$
      \lambda \hat e_i = \hat \alpha_i u + \hat \alpha_i{}^j \hat e_j
\;,
$$
with $(\hat \alpha_i )= (\hat \alpha_1, 0,\ldots, 0)$. But then
the space $\Span \{ \hat e_2,\ldots\}$ is a Riemannian subspace
of $E$ invariant under $\lambda$, which leads to a
contradiction unless $\{e_i\}$ contains only one element, and
then $\dim E=3$. This shows that $\{e_1\}^\perp$ is a
two-dimensional Lorentzian space containing $u$. We can choose
an ON basis $\{e_0,e_2\}$ of $\{e_1\}^\perp$, with $e_0$
timelike, so that $u=e_0+e_2$. The equation $\lambda u = \sigma
u $, where $\sigma\in\R$ is the eigenvalue, gives
$$\lambda u =  (\lambda ^\mu{}_0+ \lambda^\mu{}_2) e_\mu = \sigma (e_0+e_2)
\;.
$$
Equivalently, keeping in mind
$\lambda_{\mu\nu}=-\lambda_{\nu\mu}$,
$$
 \lambda^0{}_2=\sigma=\lambda^2{}_0\;, \quad \underbrace{\lambda^1{}_0{}}_{=:a}+\lambda^1{}_2=0
 \;.
$$
So
 \bel{can2x}
 \lambda^\mu{}_\nu=\left(\begin{tabular}{ccc}
               $0$ & $a $& $\sigma$
                \cr $ a$ & $0$ & $-a$
                \cr $\sigma$
                    & $a$ & $0$ \cr
              \end{tabular}
              \right)
              \ \Longleftrightarrow\
 \lambda_{\mu\nu}=\left(\begin{tabular}{ccc}
               $0$ & $-a $& $-\sigma$
                \cr $ a$ & $0$ & $-a$
                \cr $\sigma$
                    & $a$ & $0$ \cr
              \end{tabular}
              \right) \;. \ee
Calculating $\det (\lambda -\sigma\, \id)$, one finds that
$\lambda$ has  both $\sigma $ and $-\sigma$ as eigenvalues,
which at this stage is consistent only if $\sigma$ vanishes. If
$a=0$ we obtain a contradiction with the fact that $\lambda|_E$
is non-trivial, so \eq{Xfcase2} holds with $a\ne 0$, and the
result is established.
\end{proof}

\medskip

We wish, now to  show that~\cite[Theorem~1.1]{Chstatic} remains
valid in higher dimensions, under the following proviso: For
consistency of notation with the remainder of this work, let us
denote by $\hyp$ the manifold $\Sigma$ there. One then needs to
assume that the doubling of $\hyp$ across all non-degenerate
components of its boundary, and compactification of all
asymptotically flat regions except one, leads to a manifold of
positive energy type, as defined in~\cite[Section~1.1]{ChCo}.
Under this condition, the arguments of the proof
of~\cite[Theorem~1.1]{Chstatic} go through without
modifications except for the proof that there are no zeros of
$X$ on degenerate components of $\pohyp =\ohyp \setminus \hyp$.
In~\cite{Chstatic} such zeros were ``excluded" by the incorrect
Theorem~3.1 there. To take care of this, recall that it is
assumed in~\cite[Theorem~1.1]{Chstatic} that a vacuum
space-time $(\mcM,\fourg)$ has a hypersurface-orthogonal
Killing vector $\changedX$ which is timelike along a spacelike
hypersurface $\hyp$. Further, it is assumed that $\changedX$
vanishes on the boundary $\pohyp$, which is supposed there to
be a compact two-dimensional topological manifold, and which we
allow in this work to be any compact topological manifold of
co-dimension two in $\mcM$.  It is shown
in~\cite[Section~5.2]{ChCo} that the set, say $\mcE$, where
$\fourg(\changedX,\changedX)$ vanishes, is foliated by locally
totally geodesic null hypersurfaces, away from the points where
$\changedX$ vanishes. Hence each leaf of $\mcE$ is smooth on an
open dense set, so $\partial\hyp$ is smooth on the open dense
subset of $\partial \hyp$ consisting of  points at which
$\changedX$ does not vanish. Note that $\mcE$ might fail to be
embedded in general, but this is irrelevant for the proof here
because $\partial \hyp$ is a compact embedded topological
manifold by hypothesis. In vacuum, on every smooth leaf of
$\mcE$, and hence on every smooth component of $\partial \hyp$,
the surface gravity $\kappa$ is constant (see, e.g.,
\cite[Theorem~2.1]{RaczWald2}). It follows that the problem
with the incorrect~\cite[Theorem~3.1]{Chstatic} is solved by
the following:

\begin{Proposition}\label{Ponce2x}
Let $(\mcM,\fourg)$ be an $(n+1)$--dimensional Lorentzian
manifold with Killing vector field $\changedX$, and suppose
that
\bel{boundeqOm}
 \Omega:=\partial\{p\in \mcM \ | \
\fourg(\changedX,\changedX)<0\}
 \;.
  \ee
is a topological hypersurface. Assume that
\begin{enumerate}
 \item either $\changedX$ is hypersurface-orthogonal and
     $\Omega$ has vanishing surface gravity wherever
     defined,
 \item or $\Omega$ is differentiable.
\end{enumerate}
 Then $\changedX$
has no zeros on ${\Omega}$.
\end{Proposition}

\proof     Suppose, first, that $\changedX$ is of the form
\eq{Xfcase2} in a geodesically convex neighborhood $\mcU$ of
$p$ globally coordinatised by normal coordinates. This,
together with elementary properties of normal coordinates,
implies
\bel{Ynorm2} \fourg(\changedX,\changedX) = a^2(x^0-x^2)^2+
\sum_{i=1}^m \beta_i^2\left((x^{2i+1})^2+(x^{2i+2})^2\right) +
 O(|x|^4) \;,
\ee
where $|x|^2=(x^0)^2+\ldots+(x^n)^2$. It follows from
\eq{Xfcase2} that $\changedX$ is tangent to the two
hypersurfaces
$$
 \mcN^\pm=\{x^0=x^2\;,\ \pm x^2>0\}
 \;,
$$
non-vanishing there.

Consider any point $q\in \Omega$ at which $\changedX$ does not
vanish. As shown in~\cite{ChCo}, the hypersurface $\Omega$ is
smooth near $q$, and any geodesic $\gamma$ initially normal to
$\changedX_q$ stays on $\Omega$, except perhaps when it reaches
a point at which $\changedX$ vanishes.

So, suppose that $\gamma$ is such a geodesic from $q\in \Omega$
to $p$, with $p$ being the first point on $\gamma$ at which
$\changedX$ vanishes. If $\dot x^0\ne \dot x^2$ at $p$,
\eq{Ynorm2} shows that $\changedX$ is spacelike  along $\gamma$
near and away from $p$, contradicting the fact that $\changedX$
is null on $\Omega$. We conclude that $\dot \gamma$ is tangent
at $p$ to the hypersurface $\{x^0=x^2\}$, but then $\gamma\cap
\mcU$is included in $\{x^0=x^2\}$. Consequently
\bel{Ominc} \Omega\cap \mcU \subset \{x^0=x^2\} \;. \ee
Since $\Omega$ is a topological hypersurface by hypothesis, we
obtain that
\bel{Omesubs} \Omega \cap \mcU = \{x^0=x^2\} \;. \ee
(In particular $\Omega$ is smooth near $p$.)

In the case where $\changedX$ is not necessarily hypersurface
orthogonal, but we assume a priori that $\Omega$ is
differentiable, the argument is somewhat similar, with a weaker
conclusion: Let $\gamma\subset \Omega$ be any differentiable
curve, then we must have $\dot x^0=\dot x^2$ at $p$. Since
$\Omega$ is a hypersurface, this implies that
\bel{Omesubs2} T_p\Omega=T_p\{x^0=x^2\}
 \;.
\ee
So, while \eq{Omesubs} does not necessarily hold, the tangent
spaces coincide at $p$ in both cases.

Consider, now any differentiable curve $\sigma$ through $p$ on
which $\dot x^0\ne \dot x^2\ne 0$ at $p$. \Eq{Ynorm2} shows
that on $\sigma$ the Killing vector $\changedX$ is spacelike
near and away from $p$. By \eq{Omesubs2} such curves are
transverse to $\Omega$, which shows  that there exist points
arbitrarily close to $\Omega$ at which $\changedX$ is
\emph{spacelike} on both sides of $\Omega$. This contradicts
\eq{boundeqOm}, and shows that \eq{Xfcase2} cannot arise under
our hypotheses.

It remains to analyze Killing vectors of the form \eq{Xfcase}.
In this case
\bel{Ynorm2x} \fourg(\changedX,\changedX) =
\beta_0^2\left(-(x^0)^2+(x^1)^2\right)+ \sum_{i=1}^m
\beta_i^2\left((x^{2i})^2+(x^{2i+1})^2\right) +
 O(|x|^4) \;.
\ee
Suppose, first, that $\beta_0=0$. Then $\Ker
\lambda=\Span\{\partial_0,\partial_1\}|_p$. Now, because the
flow of a Killing vector maps geodesics to geodesics,
$\changedX$ vanishes on every geodesic $\gamma$ with
$\gamma(0)=p$ such that $\dot \gamma (0)\in \Ker \lambda$. So
$\changedX$ vanishes throughout the timelike hypersurface
$\{x^2=\ldots=x^n=0\}$. At every point $q$ of this
hypersurface, in adapted normal coordinates centered a $q$ the
tensor $\nabla_c \changedX_d|_q$ takes the form \eq{can} with
$\beta_0=0$. This implies that $\changedX$ is spacelike or
vanishing throughout a neighborhood of $p$, so $\beta_0=0$
cannot occur.

Now, if $\Omega$ is differentiable at $p$, an argument very
similar to the one above shows that
$$
 T_p\Omega \subset E_+\cup E_-\;, \quad \mbox{where} \ E_\pm:=\{\dot x^0=\pm \dot x^1\}
 \;.
$$
So either $T_p\Omega=E_+$ or $T_p\Omega=E_-$. But, the curves
with $\dot x^0=  \dot x^1/2$ at $p$  are transverse both to
$E_-$ and to $E_+$, with $\changedX$  spacelike on those curves
near and away from $p$ on both sides of $\mcE_\pm$,
contradicting the definition of $\Omega$. So, under the
assumption of differentiability of $\Omega$ the proof is
complete.

Assuming, next, that $\changedX$ is hypersurface-orthogonal, we
claim that $\beta_i=0$. Indeed, let $X^\flat$ be the field of
one-forms defined as $X^\flat=\fourg(X,\cdot)$. Then
\beaa
X^\flat & = &   \beta_0(x^0 dx^1-x^1dx^0)
 + \sum_{i=1}^m
\beta_i(x^{2i} dx^{2i+1}- x^{2i+1} dx^{2i}) +
 O(|x|^{3/2}) \;,
 \\
 dX^\flat & = &   2\beta_0 \, dx^0 \wedge dx^1
 + \sum_{i=1}^m 2
\beta_i\, dx^{2i}\wedge dx^{2i+1} + O(|x|^{3/2}) \;,
\eeaa
and the staticity condition $X^\flat \wedge dX^\flat =0$ gives
$\beta_i=0$, $i=1,\ldots,m$.

Arguments similar to the ones already given show now that
$$
 \Omega\cap\mcU \cap \{x^2=\ldots =x^n=0\;, \ x^0=\pm x^1\}\ne \emptyset
 \;. $$
Next, from \eq{Ynorm2x} we have
$$
 d\left(\fourg(\changedX,\changedX)\right) =
2\beta_0^2\left(-x^0 dx^0+x^1dx^1\right)+ 2\sum_{i=1}^m
\beta_i^2\left(x^{2i}dx^{2i}+x^{2i+1}dx^{2i+1}\right) +
 O(|x|^3) \;,
$$
and recall that this vanishes on $\Omega$ wherever
 $\Omega$ is
differentiable, by definition of degeneracy. But on
$S:=\{x^2=\ldots =x^n=0\;, \ x^0=\pm x^1\}$, with $|x|$
sufficiently small, we clearly have $
 d\left(\fourg(\changedX,\changedX)\right) \ne 0$. If points on
 $S$ are differentiability points of $\Omega$ we are done;
 otherwise, notice that   $
 d\left(\fourg(\changedX,\changedX)\right) \ne 0$ on a
 space-time neighborhood of $S\cap \{0<|x|<\epsilon\}$ for some
 $\epsilon>0$, and since differentiability points are dense on
 $\Omega$ the horizon cannot be degenerate.
\qed

\bigskip

Proposition~\ref{PGhani} allows us also to solve a question
concerning the codimension of zero-sets of Killing vectors
within null hypersurfaces, that arose
in~\cite[Section~5]{ChCo}:

\begin{Proposition}
 \label{Pcodim}
Let $X$ be a Killing vector. Suppose that $X$ vanishes at $p$.
Then the intersection of the zero-set of $X$ with a null
hypersurface $\mcN$ is, near $p$, a smooth submanifold of
$\mcN$ with $\mcN$--codimension at least two, unless $T\mcN$
contains a null generator on which $X$ vanishes, or is tangent
to it.
\end{Proposition}

\proof
Suppose, first, that near $p$ the Killing vector $X$ takes the
form \eq{Xfcase}.  If $\beta_0\ne 0$ and if at least one
$\beta_i=0$, with $i\ge 1$, is non-zero, then $X$ vanishes on a
smooth submanifold through $p$ of codimension larger than or
equal to four, and the result is straightforward. If
$\beta_0=0$, the result follows from the fact that the
codimension the zero set of $X$ in $\mcM$ equals that in
$\mcN$. Otherwise only $\beta_0$ is different from zero, and
the zero-set of $X$ through $p$ is a smooth spacelike
submanifold $S$ of co-dimension two. A straightforward
examination of the tangent planes at $p$ shows that the
intersection with any null-hypersurface $\mcN$ is  a set of
co-dimension at least two unless the null tangent plane of
$\mcN$ at $p$ contains one of the null normals to $S$. But then
the corresponding generator of $\mcN$ through $p$ will contain,
at least near $p$,  a null orbit of $X$ accumulating at $p$.
The analysis of \eq{Xfcase2} is similar.
\qed

\section{Concluding remarks}
 \label{SCr}

As discussed in more detail in~\cite{ChCo}, event horizons in
well behaved stationary asymptotically flat space-times are
smooth hypersurfaces. The key to the proof of this fact
is~\cite[Theorem~6.18]{ChDGH}, with a purely local proof except
for the requirement that the conclusions of the area theorem
hold. So any set of  global conditions ensuring the validity of
the area theorem imply the result. Now, smoothness of the event
horizon is needed to prove the existence of a supplementary
isometry in the space-time, via the so-called \emph{rigidity
theorem}~\cite{Ha1,HIW,VinceJimHigh}. While it is clear that
some version of this statement remains correct for
$KK$--asymptotically flat space-times, we have not investigated
this issue any further since our main results here assume  more
Killing vectors than provided by the rigidity theorem. Under
the hypotheses of Theorem~\ref{TdocA}, smoothness of the event
horizon follows from the locally totally geodesic character of
leaves of the zero-level set of the area function $W$,
see~\cite[Corollary~5.13]{ChCo}.

We note that the key elements of the uniqueness argument for non-degenerate
Kerr black holes, derived from {\regular}ity and asymptotic
flatness, are: a) simple connectedness;
  b)
smoothness of the event horizon;
 c) product structure of the domain of outer communications;
  d) the reduction of the problem to a singular harmonic map
with well understood uniqueness properties,  with e) well
understood boundary conditions.

In this paper, assuming $KK$--asymptotic flatness, we noted
that b) holds but is less essential given the number of Killing
vectors assumed; we proved c); we pointed out a version of a)
sufficient to define the twist potentials, and to prove
positivity of the area function. All this establishes d).
Theorem~\ref{TdocA}, perhaps the most involved result here, provides an
essential step towards e). However a complete proof, that the
resulting reduced equations satisfy the right boundary
conditions at $\mcA\cup\partial\doc$ for uniqueness, has to be
carried out yet, both for non-degenerate and degenerate
horizons. (Recall that the question of boundary conditions at
degenerate horizons is open even with $n=3$). We are hoping to
return to at least some of those issues in a near future.

\bigskip

{\noindent \sc Acknowledgements} I am grateful to A.~Zeghib for
pointing out Proposition~\ref{PGhani} and its proof, and to
J.~Lopes~Costa for comments about a previous version of this
paper.

\medskip

\backmatter
\bibliographystyle {smfplain}
\bibliography{
../references/newbiblio,%
../references/newbib,%
../references/reffile,%
../references/bibl,%
../references/Energy,%
../references/hip_bib,%
../references/netbiblio,../references/addon}
\end {document}